\documentclass[11pt]{article}
\usepackage[margin=1in]{geometry}

\usepackage{amsmath,amsfonts,amssymb,amsthm}
\usepackage{chngcntr}
\usepackage{mathtools}
\usepackage{bm}
\usepackage{natbib}
\usepackage{color}
\usepackage[table]{xcolor}
\usepackage{graphicx, float}
\usepackage{subfig}
\usepackage{setspace}
\usepackage[inline]{enumitem}
\usepackage{xargs, xstring}
\usepackage{stackengine}
\usepackage{booktabs}
\usepackage{multirow}

\usepackage{longtable}
\usepackage{array}
\newcolumntype{L}[1]{>{\raggedright\let\newline\\\arraybackslash\hspace{0pt}}m{#1}}
\newcolumntype{C}[1]{>{\centering\let\newline\\\arraybackslash\hspace{0pt}}m{#1}}
\newcolumntype{R}[1]{>{\raggedleft\let\newline\\\arraybackslash\hspace{0pt}}m{#1}}

\usepackage[toc,page,header]{appendix}
\usepackage{minitoc}

\usepackage[compact]{titlesec}
\titleformat{\section}
{\bfseries\centering}{\thesection.}{1em}{}
\titleformat{\subsection}
{\normalfont\normalsize\bfseries}{\thesubsection}{1em}{}
\titleformat{\paragraph}[runin]
{\normalfont\normalsize\itshape}{\theparagraph}{1em}{}

\usepackage{setspace}

\theoremstyle{plain}
\newtheorem{lemma}{Lemma}
\newtheorem{proposition}{Proposition}

\newtheorem{assumption}{Assumption}
\newtheorem{theorem}{Theorem}

\let\origtheassumption\theassumption

\newcommandx{\cluster}[1][1=i]{Q_{#1}}
\newcommand{\onecov}{X}
\newcommandx{\covs}[1][1={ij}]{\onecov_{#1}}
\newcommandx{\covsu}[0][]{\bm \onecov}
\newcommandx{\covsr}[1][1={}]{\MakeLowercase{\onecov}_{#1}}
\newcommand{\onecovs}{V}  
\newcommand{\onecovc}{V}  
\newcommandx{\covss}[1][1={ij}]{\onecovs_{#1}}
\newcommandx{\covsc}[1][1=j]{\onecovc^c_{#1}}
\newcommandx{\trt}[1][1={ij}]{Z_{#1}}
\newcommandx{\trtc}[1][1=j]{Z^c_{#1}}
\newcommandx{\trtcbm}[0][]{{\bm Z}^c}

\newcommandx{\trtr}[1][1={}]{z_{#1}}
\newcommandx{\out}[1][1={ij}]{Y_{#1}}
\newcommandx{\outr}[1][1={}]{y}
\newcommandx{\enrol}[1][1={ij}]{R_{#1}}
\newcommandx{\enrolr}[1][1={}]{r_{#1}}
\newcommandx{\stratum}[1][1={ij}]{S_{#1}}
\newcommandx{\penrol}[2][1=\trtr,2=\covsr]{e_{\enrol[]}(#2, #1)}
\newcommandx{\wps}[2][1=\covsr,2={}]{e^{#2}(#1)}
\newcommand{\wpS}{e(\covs[])}
\newcommandx{\wpsonepar}{\alpha}
\newcommandx{\wpspar}{{ \wpsonepar}}
\newcommandx{\wpsmod}[2][1=\covs, 2=\wpspar]{e(#1; #2)}

\newcommandx{\kappamod}[2][1=\covs, 2=\wpspar]{$\kappa$(#1, #2)}
\newcommandx{\weight}[2][1={ij},2={}]{{w}^{#2}_{#1}}
\newcommandx{\effect}[0][]{\tau}
\newcommandx{\data}[1][1=j]{\mathcal{O}_{#1}}
\newcommandx{\obsdata}[2][1=j,2=T]{\IfEqCase{#2}{{T}{O} {F}{o}}_{#1}}
\newcommand{\E}{\mathrm{E}}
\newcommand{\prob}{\mathrm{P}}
\newcommand{\independent}{\perp\!\!\!\perp}
\newcommand{\indep}{\perp\!\!\!\perp}

\usepackage{varioref}
\usepackage{xr-hyper}
\RequirePackage[citecolor=black,colorlinks=true,linkcolor=black]{hyperref}
\usepackage[capitalise,noabbrev]{cleveref}
\crefformat{equation}{(#2#1#3)}
\crefformat{assumption}{Assumption #1}

\newcommand{\usetitle}{Addressing selection bias in cluster randomized experiments via weighting}
\newcommand{\useauthor}{
Georgia Papadogeorgou$^1$, Bo Liu$^2$, Fan Li$^3$, Fan Li$^2$}
\newcommand{\useaffiliation}{
$^1$Department of Statistics, University of Florida, USA \\
$^2$Department of Statistical Science, Duke University, USA \\
$^3$ Department of Biostatistics, Yale University, USA
}
\newcommand{\useauthorshort}{Georgia Papadogeorgou, Bo Liu, Fan Li, Fan Li}

\newcommand\numberthis{\addtocounter{equation}{1}\tag{\theequation}}

\begin{document}

\setstretch{1.5}

\begin{center}
{\bf \Large \usetitle} \\[10pt]
{ \large \useauthor} \\[10pt]
\setstretch{1.2}
{\small \useaffiliation}
\end{center}

\begin{abstract}
In cluster randomized experiments, individuals are often recruited after the cluster treatment assignment, and data are typically only available for the recruited sample. Post-randomization recruitment can lead to selection bias, inducing systematic differences between the overall and the recruited populations, and between the recruited intervention and control arms. In this setting, we define causal estimands for the overall and the recruited populations. 
We prove, under the assumption of ignorable recruitment, that the average treatment effect on the recruited population can be consistently estimated from the recruited sample using inverse probability weighting. Generally we cannot identify the average treatment effect on the overall population. Nonetheless, we show, via a principal stratification formulation, that one can use weighting of the recruited sample to identify treatment effects on two meaningful subpopulations of the overall population: individuals who would be recruited into the study regardless of the assignment, and individuals who would be recruited into the study under treatment but not under control. We develop an estimation strategy and a sensitivity analysis approach for checking the ignorable recruitment assumption.
{The proposed methods are applied to the ARTEMIS cluster randomized trial, where removing co-payment barriers increases the persistence of P2Y$_{12}$ inhibitor among the always-recruited population.}
\end{abstract}

\noindent
{\it keywords:}
Causal inference; Cluster randomized trial; Principal stratification; Sensitivity analysis; Selection bias; Working propensity score

\section{Introduction}
\label{s:intro}

Randomized experiments are the gold standard for evaluating causal effects. Randomization guarantees that the treatment arms are comparable in both measured and unmeasured baseline covariates. However, when the inclusion of an individual unit, which we generically refer to as \emph{recruitment} or \emph{enrollment} herein, into an experiment is determined after the treatment assignment, individuals can self-select whether to participate in the experiment, inducing post-randomization selection bias. This problem is prevalent in cluster randomized experiments, where individuals are typically recruited after cluster randomization and are not blinded to the assignment. 
Patients are often more willing to participate in the study if they are under a specific treatment arm, especially when that treatment has a perceived benefit. Consequently, the recruited treatment and control arms can be systematically different, breaking the initial randomization. 

{We consider post-randomization selection as a special case of the general setting of an intermediate variable lying temporally between an exposure and an outcome \citep{Rosenbaum1984consequences, Frangakis2002principal}. 
The selection bias setting due to recruitment is unique in that data on the unrecruited individuals are entirely missing, whereas in many other settings we have at least partial data on all individuals. {Post-randomization selection bias is particularly prevalent in cluster randomized experiments where individual recruitment occurs after cluster-level treatment assignment, and blinding is frequently impractical. Also known as recruitment or identification bias, this type of selection bias was recognized by \citet{puffer2003evidence} in a review of cluster randomized experiments published from 1997 to 2002. \citet{brierley2012bias} conducted a review of 24 published cluster randomized experiments in four leading medical journals and found that 40\% identified individuals after randomization and could be subject to selection bias. Similar observations have been repeatedly confirmed in systematic reviews of cluster randomized experiments, for example, by \citet{bolzern2018review} and \citet{easter2021cluster}.} 
Due to the relevance of selection bias in the cluster randomized setting, we address a fundamental question---that is, \emph{whether one can estimate meaningful causal estimands on the overall population based on the recruited sample alone}.
{Our setting is different from the challenges arising in randomized experiments due to selective drop-out when the treatment assignment is not blinded or it can be inferred based on potential side effects \citep{little1996intent}, where data are typically available on the whole population.}

{Post-randomization selection has important implications in the analysis of cluster randomized experiments \citep{li2022clarifying}, and if ignored, often leads to ambiguous causal estimands and estimation bias. First, the recruited sample is generally not a simple random sample of the overall population; therefore, it is important to differentiate between the causal estimands defined on the recruited population and the overall population. Although previous simulations have recommended regression adjustment to address post-randomization selection bias in cluster randomized experiments \citep{leyrat2013propensity,leyrat2014propensity}, they have restricted to an overly strong homogeneous treatment effect assumption that masks the importance of target population and estimands. 
Second, in the presence of post-randomization selection, one generally cannot estimate well-defined causal estimands on the recruited population without additional assumptions \citep{wang2024model,schochet2023estimating}. 
For example, \citet{wang2024model} considered a scenario where the recruited cluster size depends on the treatment, cluster-level covariates as well as the underlying cluster population size. However, they did not address a likely scenario where the recruitment process further depends on individual-level characteristics. Finally, current methodology suggests that it is necessary to obtain additional data on at least a portion of the non-recruited individuals to validly estimate causal effects on the overall population \citep{li2022note}. However, such additional data are usually not directly available in a given experiment.} 


In this article, we clarify the causal estimands, contribute novel identification results, and develop weighting methods that leverage individual-level covariates to address post-randomization selection bias in cluster randomized experiments, without assuming homogeneous treatment effects nor additional information among the non-recruited. 
We show that, under an ignorability and a monotonicity assumption on the recruitment mechanism, one can nonparametrically identify, using only the recruited sample, the average treatment effect on (i) the recruited population, and (ii) two interpretable subpopulations of the overall population. 
The central tool is weighting by the {\it working propensity score}, which is the conditional probability of being in the treatment arm among the recruited. 
Recognizing that the ignorable recruitment assumption may not always hold, we then develop a new sensitivity analysis method to assess the impact under violations of this assumption.

{The proposed methods are motivated by the ARTEMIS cluster randomized trial \cite[the Affordability and Real-World Antiplatelet Treatment Effectiveness After Myocardial Infarction Study;][]{wang2019artemis}. The ARTEMIS trial aims at evaluating whether removing co-payment barriers increases the persistence of P2Y$_{12}$ inhibitor---a common drug for patients who had experienced myocardial infarction---and lowers risk of major adverse cardiovascular events. The intervention is randomized at the hospital level, {and patients were recruited in the study with knowledge of the hospital-level treatment assignment. Only the intervention hospitals offered vouchers for P2Y$_{12}$ to the enrolled patients with no out-of-pocket cost.} Because the intervention is an obvious incentive to the patients, enrolling patients in the control hospitals was more challenging than the intervention hospitals, leading to potential selection bias due to recruitment.
Using our methods, we identify a positive causal effect of removing co-payment barriers on persistence of P2Y$_{12}$ inhibitor in the patient subpopulation that would be recruited under either treatment arm.
Furthermore, in ARTEMIS, since individuals in the treated clusters were readily recruited, the patient population in the treated clusters is likely representative of the overall population. We report an increase in medication persistence on this treated population, which is suggestive of a potential positive effect of removing co-payment barriers on the overall population.
} 

\section{Notation and Causal Estimands}
\label{sec:estimands}

Consider a cluster randomized experiment which consists of $J$ clusters drawn from a super-population of clusters, among which $m$ clusters are randomized to the intervention arm, denoted by $Z_j^c=1$, and the remaining clusters to the control arm, denoted by $Z_j^c=0$. The superscript `$c$' indicates a cluster-level variable.
{Even though treatment randomization is considered the gold-standard for drawing causal inferences, data from cluster randomized experiments can suffer from selection bias stemming from individual decisions on cluster membership and enrollment after the cluster-level treatment assignment \citep{schochet2023estimating}.}

Whether selection bias due to cluster membership takes place depends on whether the treatment assignment affects the individual choice of cluster.
For example, in health care of rare diseases, patients might choose a hospital that is specialized in treating their conditions or offers a new experimental treatment that is perceived as beneficial. {Such decision making process on individual cluster membership would render treated and control clusters imbalanced in their composition and would likely lead to a larger load of patients in treated hospitals.} We do not consider this scenario and refer to \cite{schochet2023estimating} for more details.  
Instead, we focus on the scenario where individuals do not self-select into a cluster based on the cluster-level treatment assignment. For example, individuals in a classroom or village are usually fixed at the time of randomization and would remain in that cluster throughout the study. Another example is health care of common diseases such as heart disease or diabetes; patients usually choose a hospital based on distance or insurance and would not change hospitals. This is the case in our study (\cref{sec:application}). We formalize this scenario in Assumption \ref{ass:choosing_cluster}. 

\begin{assumption} \label{ass:choosing_cluster} (\emph{Constant cluster membership})
Let $\mathcal J, \mathcal{J'}$ denote two sets of clusters that could comprise the cluster randomized experiment, and $\bm z \in \{0, 1\}^{|\mathcal J|}, \bm z' \in \{0, 1\}^{|\mathcal J'|}$ two hypothetical cluster treatment assignments in each composition, respectively. Let $\cluster[](\mathcal J, \bm z)$ denote the cluster that a randomly chosen individual would belong to under cluster composition $\mathcal J$ and treatment assignment $\bm z$. We assume that $\cluster[](\mathcal J, \bm z) = \cluster[](\mathcal J', \bm z')$.
\end{assumption}

Under \cref{ass:choosing_cluster}, if we conceive that individuals form a super-population and they arrive at the super-population of clusters under some random process, then this process is constant with respect to the study, and the individuals in the cluster randomized experiment can be considered fixed. We refer to all individuals, recruited or not, across the clusters in the experiment as the {\it overall population}. \cref{ass:choosing_cluster} also allows us to use a double-index notation: `$ij$' denotes individual $i$ in cluster $j$ with the total number of individuals $N_j$. Under cluster randomization, we also have $\trt = \trtc$. We use $N =\sum_j N_j$ to denote the number of individuals in the overall population.

{The second source of selection bias in cluster randomized experiments arises from selective enrollment or recruitment into the study of individuals in the overall population that are eligible for the experiment.
Selective recruitment is common in cluster randomized experiments in the healthcare delivery context across disease conditions, as suggested by a series of reviews \citep{puffer2003evidence,brierley2012bias,bolzern2018review}, and has been found to be particularly common in cluster randomized studies evaluating individual-level interventions \citep{easter2021cluster}. Mechanistically, after clusters are randomized, individuals from treated clusters may be more likely to give consent to join the study when the intervention is perceived to be beneficial, or recruiters from clusters may be more motivated to enroll individuals based on their foreknowledge of cluster assignment.} In either case, the subpopulation recruited into the study (denoted by $\enrol=1$) is referred to as the \emph{recruited population}, which is different from the overall population as long as some individuals are unrecruited (denoted by $\enrol=0$). We denote the number of recruited individuals in cluster $j$ by $n_j \leq N_j$, and the total number of recruited individuals in the study by $n = \sum_j n_j$.

Let $\covss$ and $\covsc$ denote the covariates for individual $i$ in cluster $j$, and for cluster $j$, respectively, and write $X_{ij}=(\covss, \covsc)$ as the collection of covariates for individual $i$. Under the cluster-level Stable Unit Treatment Value Assumption, there are two individual-level potential outcomes $\out(0)$ and $\out(1)$ as well as two potential recruitment statuses, $\enrol(0)$ and $\enrol(1)$. We only observe the covariates and the potential outcomes corresponding to the observed assignment of the recruited individuals, denoted as $\out=\out(\trt)$ for those with $\enrol=\enrol(\trt) = 1$. In what follows, we always use the notation `$\sum_{ij}$' to denote a sum over the {\it recruited individuals} across all clusters.


We define the \emph{average treatment effect on the overall population} as
\begin{align*} 
\tau^O=\E\left\{\out(1)-\out(0) \right\}.
\end{align*} 
We also define the \emph{average treatment effect on the recruited population} and its counterpart specific to treatment status $z \in \{0, 1\}$ as
\begin{align*} 
\tau^R=\E\left\{\out(1)-\out(0)\mid \enrol=1\right\}; \quad \tau^R_{\trtr} =
\E\left\{\out(1)-\out(0)\mid \trt = \trtr, \enrol=1\right\}, \end{align*}
respectively. The expectation is over the cluster super-population, from which the cluster information $\{V_j^c, N_j, \{V_{ij}, R_{ij}(z), Y_{ij}(z)\}_{i=1}^{N_j}\}$ are independent and identically distributed draws. 
The estimand on the overall population $\tau^O$ and its counterparts on certain subpopulations of the overall population are usually the intended target estimands. However, researchers often resort to estimate $\tau^R$ because usually only data on the recruited sample are available. When the recruited sample is a simple random sample of the overall population or the treatment effect is homogeneous across all individuals, $\tau^O$ is equal to $\tau^R$, but not so otherwise.

We now introduce causal estimands on meaningful subpopulations of the overall population via a principal stratification formulation \citep{Frangakis2002principal}. 
We classify the overall population by the joint individual potential recruitment statuses under both assignments: $\stratum=(\enrol(0),\enrol(1))$; $\stratum$ is called a principal stratum. Due to the fundamental problem of causal inference, the individual principal stratum membership is not observable.
We name the four principal strata as:  
$\stratum= (1,1) \equiv a$, \textit{a}lways-recruited, individuals who would be recruited regardless of the assignment;
$\stratum=(0,0)\equiv n$, \textit{n}ever-recruited, individuals who would not be recruited regardless of the assignment;
$\stratum=(0,1)\equiv c$, in\textit{c}entivized-recruited, individuals who would be recruited under intervention but not under control;
and $\stratum=(1,0)\equiv d$, \textit{d}isincentivized-recruited, individuals who would be recruited under control but not under intervention. In the noncompliance literature \citep{Angrist1996identification}, the latter two groups are often referred to as compliers and defiers, respectively.
By construction, principal stratum membership does not change by treatment assignment; therefore, we can define causal effects within each principal stratum, for example
\begin{equation}
\tau^O_{s} = \E\left\{ \out(1) - \out(0) \mid \stratum = s\right\}, 
\end{equation}
for $s\in  \{a, n, c, d\}$.
If $\pi_{s}$ represents the proportion of principal stratum $s$ in the overall population, then $\tau^O = \sum_s \pi_s\tau^O_{s}$. We also define causal effects on unions of principal strata,  $\tau^O_{a,c} = \E\left\{\out(1) - \out(0) \mid \stratum \in \{a, c\}\right\}$, which is equal to the weighted average of $\tau^O_a$ and $\tau^O_c$ with weights $\pi_a / (\pi_a + \pi_c)$ and $\pi_c / (\pi_a + \pi_c)$, respectively.

The relevance of an estimand for policy-making critically depends on the scientific context. In general, estimands on the recruited population can be hard to interpret since this population corresponds to an imbalanced mixture of always-, incentivized-, and disincentivized-recruited individuals \citep{li2022clarifying}. However, our study in \cref{sec:application} represents a situation where estimands on a subpopulation of the recruited population can be informative of general policy. In contrast, the causal effects defined over unions of principal strata weigh the strata populations according to their prevalence in the overall population, and therefore provide a more interpretable representation of the population. That said, since principal stratum membership is always unobserved, it is not immediately clear how to use strata-specific estimands for policy-making. In \cref{sec:application} we illustrate how one may design policy to target various principal strata by investigating their baseline covariate distributions.

\section{Nonparametric Identification of Causal Estimands} \label{sec:identification}
\subsection{Identification Assumptions} \label{sec:assumptions}
\cite{li2022clarifying} showed that a simple difference in means estimator using the recruited sample is generally biased for both $\tau^O$ and $\tau^R$. In this section, we show how to nonparametrically identify the above causal estimands \emph{based on the recruited sample alone}. We first need to formalize the random assignment of the clusters.

\begin{assumption} \label{ass:randomization} (\emph{Cluster randomization}). Let $\trtc[] = (\trtc[1], \trtc[2], \dots, \trtc[J])$ denote the assignment of the clusters. Then,
$ \Pr \big( \trtc \mid \out(0), \out(1), \covss, \covsc \text{ for all } {i,j} \big)= \Pr(\trtc).$
\end{assumption}


Under Assumptions \ref{ass:choosing_cluster} and \ref{ass:randomization}, we can prove that cluster randomization implies individual randomization, as stated in the next lemma (the proof is in Supplement 
B).

\begin{lemma} If Assumptions \ref{ass:choosing_cluster} and \ref{ass:randomization} hold, then the individual-level treatment assignment on the overall population is as-if randomized at the individual level, and the probability of treatment of an individual is equal to the cluster-level treatment probability, i.e.
$\Pr(\trt = 1 \mid \out(0), \out(1), \covss, \covsc \text{ for all } {i,j} \big) = \Pr(\trt = 1)= \Pr(\trtc = 1)$. 
\label{prop:overall}
\end{lemma}

Assumption \ref{ass:choosing_cluster} characterizes the composition of the overall population. 
We now specify two related assumptions that characterize the composition of the recruited population.

\edef\oldassumption{\the\numexpr\value{assumption}+1}

\setcounter{assumption}{0}
\renewcommand{\theassumption}{\oldassumption\Alph{assumption}}

\begin{assumption} (\emph{Non-differential recruitment}) The recruitment process is non-differential with respect to potential outcomes given covariates if there exists a function $\delta(x)$ such that
\begin{equation} \label{eq:delta_def}
  \delta(x)=\frac{\Pr(\enrol = 1 \mid \out(0) = \outr_0, \out(1) = \outr_1, X_{ij}=x, \trt = 1)}{\Pr(\enrol = 1 \mid \out(0) = \outr_0, \out(1) = \outr_1, X_{ij}=x, \trt = 0)},
\end{equation}
for all $x, y_0$, and $y_1$.
\label{ass:po_diffenrol}
\end{assumption}

\begin{assumption} (\emph{Ignorable recruitment}) Conditional on covariates and the treatment assignment, an individual's recruitment status is independent of the potential outcomes:
$\Pr(\enrol = 1 \mid \out(0), \out(1), \trt,  X_{ij}) = \Pr(\enrol = 1 \mid \trt,  X_{ij})$.
\label{ass:enrol_indep}
\end{assumption}

Assumption \ref{ass:po_diffenrol} states that all factors driving the recruitment {\it differently} between treated and control individuals are measured, whereas Assumption \ref{ass:enrol_indep} states that all factors driving the recruitment are measured, i.e. there is no unmeasured confounding with respect to the recruitment process.
Assumption \ref{ass:enrol_indep} implies Assumption \ref{ass:po_diffenrol}, but not vice versa. 
As we explain later, the function $\delta(x)$ is closely related to the proportion of individuals that belong to the different principal strata.
\cref{ass:po_diffenrol} is equivalent to the subset ignorability in  \cite{gaebler2022causal}: $\trt \independent \{\out(0), \out(1)\} \mid \{ X_{ij}, \enrol = 1\}$. Since the treatment assignment temporally precedes the recruitment, it is arguably hard to justify subset ignorability in our setting because it imposes the independence of the treatment assignment conditional on the post-assignment recruitment.
{The ignorability assumption in \cite{schochet2023estimating} carries the same issue in terms of the variables' temporal ordering. Meanwhile, \cref{ass:enrol_indep} is equivalent to conditional independence of the principal stratum membership and potential outcomes \citep[referred to as principal ignorability by][]{ding2017principal}: $\stratum \independent \{\out(0), \out(1)\} \mid X_{ij}$}. (The relationship between the different assumptions is formally stated and proven in Supplement B.2.) 
{For $\delta(x)$ to be well-defined, \cref{ass:po_diffenrol} is necessary. From here onwards, when we refer to $\delta(x)$, \cref{ass:po_diffenrol} is implicitly assumed to hold.}

We maintain a standard positivity assumption that ensures that the probability of recruitment of any individual in the overall population is bounded away from 0.

\let\theassumption\origtheassumption

\setcounter{assumption}{3} 
\begin{assumption}[\emph{Recruitment positivity}]
There exists $\delta > 0$ such that $\Pr(\enrol = 1 \mid \trt = z, X_{ij} = x) > \delta$ for all $(z, x)$.
\label{ass:positivity_enrollment} 
\end{assumption}

\subsection{Nonparametric Identification of Recruited and Overall Estimands} \label{sec:identify-tau}

We first introduce the concept of a \emph{working propensity score}, defined as the probability of being in the treatment group among the recruited: $\wps = \Pr(\trt = 1\mid X_{ij}=x,\enrol=1).$
Importantly, the working propensity score does {\it not} reflect the true treatment assignment mechanism \citep{rosenbaum1983central}, but is rather a one-dimensional summary of the measured covariates of the recruited individuals. Its true value depends on the conditional distribution of the recruitment process through
\begin{align*}
\wps = \frac{\Pr(\enrol=1\mid \trt = 1,X_{ij}=x)}{\Pr(\enrol=1\mid X_{ij}=x)}\Pr(\trt = 1),
\end{align*}
which is unknown and cannot be estimated without data on the un-recruited. We discuss how to estimate the working propensity score in Section \ref{sec:estimation}. 
If Assumption \ref{ass:positivity_enrollment} holds, then the working propensity score is bounded away from 0 and 1 (see Supplement 
B.4).

With the recruited sample, $\tau^R$ is identifiable using inverse probability weighting, as
\begin{align} \label{eq:iden-ATE-R}
\tau^R &= \E \left[  \frac{\trt \out}{e(\covs)} - \frac{(1 - \trt)\out}{1 - e(\covs)} \ \bigg| \ \enrol = 1 \right].
\end{align}
Here, \cref{ass:po_diffenrol} suffices in place of the stronger \cref{ass:enrol_indep}.
The proof is in Supplement 
B.5.
Note that $\tau^R$ is also identifiable using the standard outcome modeling strategy.
The identification formula for $\tau^R_z$ is similar, with the contribution of each individual multiplied by $\wps[\covs]$ for $\tau^R_1$ and by $1 - \wps[\covs]$ for $\tau^R_0$ \citep{li2018balancing}.

We next show how to identify causal effects on two meaningful subpopulations of the overall population from the recruited sample alone.
Notice that the observed cells of $(R,Z)$ consists of mixtures of principal strata. Specifically, the recruited individuals from the intervention clusters $(\trt = 1, \enrol=1)$ consist of always- and incentivized-recruited, whereas the recruited individuals from the control clusters $(\trt = 0, \enrol=1)$ consist of always- and disincentivized-recruited. 
This observation connects two estimands for the overall population to estimands for the recruited population, summarized in \cref{prop:identifiability_overall}.

\begin{theorem}
If Assumptions \ref{ass:choosing_cluster}, \ref{ass:randomization} and \ref{ass:enrol_indep} hold, we have
\begin{enumerate*}[label=(\alph*)]
\item \label{item:overall_control}
$\tau^O_{a, d}=
\tau^R_0$, and 
\item \label{item:overall_treated}
$\tau^O_{a, c}
=\tau^R_1.$
\end{enumerate*}
If \cref{ass:positivity_enrollment} also holds, the average treatment effect on the always- and disincentivized-recruited subpopulation of the overall population, $\tau^O_{a,d}$, is identifiable as
\begin{equation}
   \tau^O_{a,d} \equiv \E [\out(1)-\out(0)\mid \stratum\in \{a, d\}]= \E \left[ \frac{\trt \out\{1-e(X_{ij})\}}{e(X_{ij})} - (1 - \trt)\out \bigg | \enrol = 1 \right], \label{eq:iden-ATE-a}
\end{equation}
and the average treatment effect among the always- and incentivized-recruited subpopulation of the overall population, $\tau^O_{a,c}$, is identifiable as
\begin{equation}
   \tau^O_{a, c}\equiv \E [\out(1)-\out(0)\mid \stratum \in \{a, c\}] = \E \left[ \trt \out - \frac{(1 - \trt)\out e(X_{ij})}{1-e(X_{ij})} \bigg | \enrol = 1 \right]. \label{eq:iden-ATE-ac}
\end{equation}
\label{prop:identifiability_overall}
\end{theorem}

\cref{prop:identifiability_overall} states that the causal effect on the union of always- and disincentivized-recruited in the overall population is equal to the causal effect on the recruited control arm, and that the causal effect on the union of the always- and incentivized-recruited in the overall population is equal to the effect on the recruited treated arm. The two causal estimands on the recruited population are identifiable via a weighting scheme that was originally designed for identifying the average treatment effect for the control and the treated individuals in the standard causal literature, respectively \citep{li2018balancing}. Consequently, the corresponding overall population estimands are also identifiable. \cref{prop:identifiability_overall} becomes further interpretable under a monotonicity assumption on the enrollment status \citep{Angrist1996identification}.

\begin{assumption}\label{ass:monotonicity} (\emph{Recruitment monotonicity}) $\enrol(1)\geq \enrol(0)$ for all individuals.
\end{assumption}
Monotonicity rules out the disincentivized-recruited, so that the recruited control arm consists only of the always-recruited, and $\tau^O_{a,d} \equiv \tau^O_a$. \cref{prop:identifiability_overall} implies a weighting-based identification strategy for $\tau^O_a$: re-weight the recruited treatment individuals by $\{1-e(X_{ij})\}/e(X_{ij})$ and leave the recruited control individuals as they are.
A similar weighting strategy allows us to identify the causal effect on the union of the always- and incentivized-recruited, $\tau^O_{a,c}$, based on the recruited sample alone. 

Next, because $\tau^O_{a,c}=\nu \tau^O_{a}+(1-\nu)\tau^O_{c}$ with $\nu={\pi_a}/(\pi_a + \pi_c)$, we can identify $\tau^O_c$ as long as we can identify the relative proportions of the always- and incentivized-recruited groups. The next result shows how to identify $\nu$ and consequently $\tau^O_{c}$.

\begin{theorem}
If Assumptions \ref{ass:choosing_cluster}, \ref{ass:randomization} and \ref{ass:monotonicity} hold, the proportion of the always-recruited among the always- and incentivized-recruited individuals is identifiable as
\begin{equation}
\nu \equiv\frac{\pi_a}{\pi_a + \pi_c} =
\frac{\pi^t (1 - p^t)}{(1 - \pi^t) p^t},
\label{eq:ident_prop_alw}
\end{equation}
where $\pi^t = \Pr(\trt[] = 1)$ and $p^t = \Pr(\trt[] = 1 \mid \enrol[] = 1)$ is the probability of treatment in the overall and the recruited population, respectively. If Assumptions \ref{ass:enrol_indep} and \ref{ass:positivity_enrollment} also hold, the causal effect among the incentivized group is identifiable as 
$$\tau^O_c=\frac{\tau^O_{a,c}-\nu \tau^O_{a}}{1-\nu},$$ 
with $\tau^O_{a,c},\tau^O_{a}$ being identified from Theorem \ref{prop:identifiability_overall}.
\label{theorem:compliers}
\end{theorem}

From \cref{prop:overall}, the probability of treatment in the overall population $\pi^t$ is equal to the probability of treatment for a cluster $P(\trtc[] = 1)$, and therefore all quantities in \cref{theorem:compliers} are identifiable based on the observed data among the recruited alone. That is, Theorems \ref{prop:identifiability_overall} and \ref{theorem:compliers} show that we can identify the causal effect among the always- and incentivized-recruited subpopulations of the overall population using data {\it only} on the recruited sample. To characterize these latent subpopulations, we can calculate the weighted average of the covariates in the always- and incentivized-recruited strata using the estimated working propensity score (introduced in Section \ref{sec:estimation}). We illustrate this approach in \cref{sec:application}.

\section{Estimation via Working Propensity Score Weighting} \label{sec:estimation} 


\subsection{Causal Effect Estimators}
Given the identification formulas \cref{eq:iden-ATE-R}, \cref{eq:iden-ATE-a} and \cref{eq:iden-ATE-ac}, we propose the corresponding Haj\'ek weighting estimators for $\tau^R$, $\tau^O_a$ and $\tau^O_{a,c}$ using only the sample of recruited individuals:
\begin{equation} \label{eq:est_wt}
\widehat{\tau}=\frac{\sum_{ij} w_1(X_{ij}) \trt \out}{\sum_{ij} w_1(X_{ij}) \trt} -
              \frac{\sum_{ij} w_0(X_{ij})(1 - \trt) \out}{\sum_{ij} w_0(X_{ij}) (1 - \trt)},
\end{equation}
with the weights $\{w_0(x)=1/(1-e(x)), w_1(x)=1/e(x)\}$ for $\tau^R$, $\{w_0(x)=1, w_1(x)=(1-e(x))/e(x)\}$ for $\tau^O_a$, and $\{w_0(x)=e(x)/(1-e(x)), w_1(x)=1\}$ for $\tau^O_{a,c}$.
We estimate $\nu$ in \cref{eq:ident_prop_alw} by substituting the probabilities of treatment in the overall and recruited population with the randomization probability at the cluster level $\Pr(\trtc[] = 1)$, and the proportion of treated individuals in the recruited sample, respectively. Then, we estimate $\tau^O_c$ using the estimators $\widehat \nu$, $\widehat \tau^O_a$ and $\widehat \tau^O_{a,c}$ based on Theorem \ref{theorem:compliers}. We show that the resulting causal estimators are consistent and asymptotically normal using M-estimation \citep{van2000asymptotic}.

\begin{theorem} 
If Assumptions \ref{ass:choosing_cluster}, \ref{ass:randomization}, \ref{ass:enrol_indep}, \ref{ass:positivity_enrollment} and \ref{ass:monotonicity} 
hold, and under standard regularity conditions,
we have that
\( \displaystyle
\sqrt{J} \big\{ (\widehat{\tau}_a^O, \widehat{\tau}_{c}^O, \widehat{\tau}^R)^\top - ({\tau}_a^O, \tau_c^O, \tau^R)^\top \big\} \rightarrow N(0, \Sigma), \) as $J \rightarrow \infty$
where the form of the asymptotic covariance matrix $\Sigma$ is given in the Supplement.
\label{prop:asym_norm}
\end{theorem}


In brief, the regularity conditions outlined in the Supplement state that cluster sizes and outcomes are bounded quantities. Furthermore, the cluster size is assumed to be non-informative of the potential outcomes. Our results readily extend to the case of informative cluster sizes, and can be extended to address a treatment effect estimand that weighs each cluster equally \citep{papadogeorgou2019causal,kahan2023estimands,wang2024model}.
}

\subsection{Estimation of the working propensity score}
\label{subsec:wps_estimation}

In practice, the working propensity score $e(x)$ is usually unknown and has to be estimated. Note that the ratio of recruitment probability $\delta(x)$ defined in \cref{eq:delta_def} is related to the working propensity score $e(x)$ through
\( \wps = {\delta(x)}/\{\delta(x) + r^{-1}\}, \)
where $r =\Pr(\trtc[] = 1) / \Pr(\trtc[] = 0)$ (see Supplement 
B.4).
%
%
Under recruitment monotonicity (\cref{ass:monotonicity}), $\delta(x)$ is also related to the probability of principal stratum membership as
\[
\delta(x) = \Pr(\stratum[] = a \mid \stratum[] \in \{a, c \}, X = x)^{-1} \geq 1
\]
(see Supplement B.3).
%
Therefore, a parametric specification can be placed on $\Pr(\stratum[] = a \mid \stratum[] \in \{a, c \}, X = x)$, $\delta(x)$, or $\wps$, but it must satisfy that $\delta(x) \geq 1$.
For propensity score estimation in general, the traditional approach is to specify a logistic model on $\wps$. However, such specification is not compatible with data from experiments with recruitment bias because it does not enforce that $\delta(x) \geq 1$. {Interestingly, this result on the recruitment probability ratio does not rely on the key identifying assumptions on the recruitment (Assumptions \ref{ass:po_diffenrol} and \ref{ass:enrol_indep}) or positivity (Assumption 4), other than the fact that the function $\delta(x)$ is well-defined (\cref{ass:po_diffenrol}).}

For that reason, we choose to specify $\delta(x)$ instead of $e(x)$. Specifying $\delta(x)$ also reflects the correct temporal order of the variables since the treatment occurs temporally {\it before} recruitment.
We adopt a specification on $\Pr(\stratum[] = a \mid \stratum[] \in \{a, c \}, X = x)$ based on parameters $\wpspar$, which leads to a parametric specification on $\delta(x)$ and $e(x)$. We specify a logistic model
\(  \displaystyle \Pr(\stratum[] = a \mid \stratum[] \in \{a, c \}, X = x) = \mathrm{expit} (x^T \wpspar ) \), which implies that $\delta(x; \wpspar) = 1 + \exp ( - x^T \wpspar )$ and
\( \mathrm{logit} \{ \wpsmod[\covsr] \}= \log \{ r \delta(x; \wpspar ) \}. \)

{To estimate $\alpha$, it is important to recognize two features of the data structure. First, all individuals within the same cluster have the same treatment. 
This precludes the inclusion of cluster-level random or fixed effects in the propensity score model since the corresponding parameters would not be identifiable. Second, by definition, the working propensity score model is a marginal model with respect to clusters and not conditional on any cluster-specific intercept. Thus, a robust and computationally efficient approach for estimation is 
to maximize the pseudo-likelihood 
among the recruited sample assuming working independence:}
\( 
\text{pseudo-}\mathcal{L}(\alpha) = \prod_{ij} \wpsmod^{\trt} \left\{1 - \wpsmod\right\}^{1 - \trt},
\)
which under the specification above becomes:
\( \text{pseudo-}\mathcal{L}(\alpha) = \prod_{ij} 
\log \{ r \delta(\covs ; \wpspar ) \} ^{\trt}
\left\{1- \log \{ r \delta(x; \wpspar ) \} \right\}^{1 - \trt}.
\)
A standard optimizer can be used to find the maximizer of the pseudo-likelihood $\widehat \wpspar$. We use the function \texttt{optim} in R.
{
Since the working propensity score is primarily used to balance treated and control recruited populations rather than perfectly predict the assignment, assuming working independence during estimation does not bias the final causal effect estimator (also empirically supported by our simulation results in \cref{sec:simulation}).} 
Extending the asymptotic distributions for the causal estimators in \cref{prop:asym_norm} to the case with the estimated propensity scores can be readily achieved by modifying the estimating function to include the pseudo-likelihood contribution for each cluster. We found in simulations (\cref{sec:simulation}) that a bootstrap procedure that resamples clusters performs well for inference based on the true or the estimated working propensity scores. {An R package for estimating the working propensity score and implementing our methods is available at \url{https://github.com/gpapadog/CRTrecruit}.}

Finally, we note that alternative specifications of $\delta(x)$ may be considered, as long as they impose that $\delta(x) \geq 1$. This is a topic for future research. 


\section{Sensitivity Analysis under Non-Ignorable Recruitment} \label{sec:SA}

The ignorable recruitment assumption (\ref{ass:enrol_indep}) is central to identifying causal effects for the overall population. We develop a sensitivity analysis to assess this assumption within the Rosenbaum's bounds framework \citep{rosenbaum2002observational}. Assume there exists an unmeasured covariate $U$ that is necessary for Assumption \ref{ass:enrol_indep}, that is, $R \independent  Y(0), Y(1)  \mid U, X, Z$, but $R \not\!\independent Y(0), Y(1) \mid X, Z$. Then, consistent estimation of the causal effects requires that we use
\[
\delta^*(x, u) = \frac{\Pr(R = 1 \mid U = u, X = x, Z = 1)}{\Pr(R = 1 \mid U = u, X = x, Z = 0)},
\]
and the corresponding working propensity score $e^*(x,u)$, instead of $\delta(x)$ and $\wps$.
We consider violations to the ignorable recruitment assumption with respect to $U$ up to a factor $\Gamma$ (the sensitivity parameter) as 
\begin{equation} \label{eq:gamma_def}
\Gamma^{-1} \leq \rho(x, u) ={\delta(x)} \ / \ {\delta^*(x, u)} \leq \Gamma.
\end{equation}
A larger value of $\Gamma$ allows a greater degree of violation. To conduct a sensitivity analysis, we need to bound the chosen estimator for each fixed value of $\Gamma$. 
We focus on the Haj\'ek estimator  \cref{eq:est_wt}, which is more efficient than the Horvitz-Thompson estimator, but is technically more challenging for sensitivity analysis \citep{aronow2013interval, papadogeorgou2022causal}.

For the causal effect on the always-recruited, $\tau^O_a$, the weights of the individuals from the control clusters are equal to 1. Therefore it suffices to focus on the part of the estimator corresponding to the individuals from the treated clusters. Denote the weights for the treated individuals in \cref{eq:est_wt} for $\tau^O_a$ as $w_{1ij}^* = w_1^*(X_{ij}, U_{ij}) = \{1 - e^*(X_{ij}, U_{ij})\} / e^*(X_{ij}, U_{ij})$. The following proposition provides an algorithm to acquire the bounds of the estimator for violations of the ignorable recruitment assumption up to $\Gamma$.


\begin{proposition} \label{prop:sensitivity}
Maximizing (minimizing) $ \tau^*_1 = \displaystyle\frac{\sum_{ij} w_{1ij}^* \trt \out}{\sum_{ij} w_{1ij}^* \trt} $ under the $\Gamma-$violation in \cref{eq:gamma_def} is equivalent to solving the linear program that maximizes (minimizes)
\(
\sum_{ij} \lambda_{ij} \weight[1ij] \trt\out 
\) 
with respect to $\lambda_{ij}$ subject to three constraints:
\begin{enumerate*}[label = (\alph*)]
\item 
\( {\kappa}{\Gamma^{-1}} \leq \lambda_{ij} \leq \kappa \Gamma \),
\item 
\(
\sum_{ij} \lambda_{ij} \weight[1ij] \trt = 1,
\)
and
\item
\(
\kappa \geq 0,
\)
\end{enumerate*}
where $\weight[1ij] = w_1(X_{ij})$ is the weight of individual $i$ under treatment and working propensity score $e(\covs)$, and $\kappa$ is a parameter of the linear program.
\end{proposition}

Bounding the causal effect on the incentivized-recruited subpopulation is more complicated because it involves a non-linear optimization problem. Instead, we acquire bounds for the causal effect estimator of the union of the always- and incentivized-recruited individuals, $\widehat \tau^O_{a, c}$, using a procedure similar to the one in \cref{prop:sensitivity}. Then, the bounds for the two estimators are combined according to the formula in \cref{theorem:compliers} to form bounds for the estimator for the incentivized-recruited principal causal effect. This procedure is computationally efficient because it only requires solving two relatively simple linear programs. It leads to conservative bounds under the $\Gamma$-violation of the ignorability assumption in the sense that the resulting bounds are no narrower than the sharp bounds that would be obtained via directly optimizing the causal effects of the incentivized-recruited subpopulation. We provide additional details in Supplement 
B.8.

\section{Simulation Studies} \label{sec:simulation}

We performed simulations to study the differences between the causal estimands in the overall and recruited populations and evaluate the properties of the estimators proposed in \cref{sec:estimation}. 
We considered 36 simulation scenarios, described in \cref{tab:sim_setup}. In each scenario, we generated $J$ clusters of size 100 as the overall population. The cluster-level treatment was assigned to 50\% or 25\% of the clusters, representing balanced and imbalanced designs, respectively. 
The proportion of recruited individuals depended on the specific data generating process and varied from 21\% to 38\%.
Data under Scenario B and C had the highest and lowest proportion of recruitment, respectively. The proportion of treated individuals in the recruited population varied from 40\% to 73\%. Data under Scenarios A and B had the same prevalence of treatment within the recruited population. We considered three individual-level and two cluster-level covariates, and generated outcomes with treatment effect heterogeneity as well as a moderate intraclass correlation coefficient (equal to 0.1).
We considered two cases corresponding to moderate and strong separation in the covariates of the always- and incentivized recruited individuals (Case 1 \& 2). The true overall average treatment effect, $\tau^O$, was 3. The average treatment effect varied within each subpopulation across the 36 scenarios and across subpopulations, from 2.71 to 2.81 for the recruited, from 2.64 to 2.68 for the always-recruited, and from 2.88 to 3.07 for the incentivized-recruited. 
The average effect among the recruited varied with the treatment probability since the corresponding subpopulation changes, even though the average effect on the overall or the always-recruited populations remain constant. This illustrates that estimands on the recruited population are less interpretable, since their interpretation can be driven by the experimental design. Full details on the data generating process are given in Supplement 
C.1.

\begin{table}[!b]
    \centering
       \caption{Specifications of the data generating models for the 36 simulation scenarios. Each scenario corresponds to a different combination of principal strata prevalences, covariate differences in always and incentivized-recruited, treatment proportion, and number of clusters.}
    \begin{tabular}{lll}
    \hline
    Principal strata prevalences & Scenario A & 20\% Always, 20\% Complier, 60\% Never \\
    & Scenario B & 25\% Always, 25\% Complier, 50\% Never \\
    & Scenario C & 15\% Always, 25\% Complier, 60\% Never  \\
      \hline
     Covariate separation of always- & Case 1 & Strong \\
     and complier-recruited & Case 2 & Moderate  \\
     \hline
     Treatment proportion & Balanced & Probability of cluster being treated is 50\% \\
     & Imbalanced & Probability of cluster being treated is 25\% \\
     \hline
     Number of clusters $J$ & 200, 500, or 800 \\ \hline
        \end{tabular}
    \label{tab:sim_setup}
\end{table}

We evaluated the estimators in terms of bias, variance and coverage of 95\% confidence intervals. We considered interval estimators based on the asymptotic theory in \cref{prop:asym_norm} for the known propensity score. We also consider interval estimators based on normal approximation with a nonparametric bootstrap procedure for the estimated propensity score. {This bootstrap procedure resamples clusters while holding the proportion of treated clusters fixed to maintain the overall prevalence of treatment. When a cluster is sampled, all individuals in the cluster are included in the bootstrap sample.} 
 
\begin{figure}
\centering
\includegraphics[width=\textwidth]{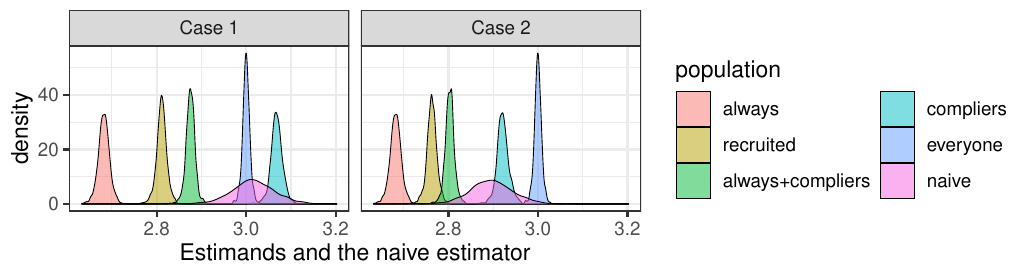}
\caption{Distribution of sample causal effect among different populations and estimated na\"ive difference of mean outcomes across 500 simulated data sets and number of clusters $J \in \{200, 500, 800\}$ in Scenario B under a balanced design in Case 1 (left) and Case 2 (right).}
\label{fig:naive_estimator_scenB}
\end{figure}

In \cref{fig:naive_estimator_scenB}, we show the true average causal effect among the different subpopulations and the estimates of the na\"ive estimator, under Scenario B with a balanced design. The na\"ive estimator, which is defined as the difference of mean outcomes among the treated recruited and the control recruited individuals, does not estimate a causal effect over any of the populations of interest across all scenarios in our simulations (see also 
Figure S.1
of the Supplement).

\begin{figure}
\centering
\includegraphics[width=0.9\textwidth]{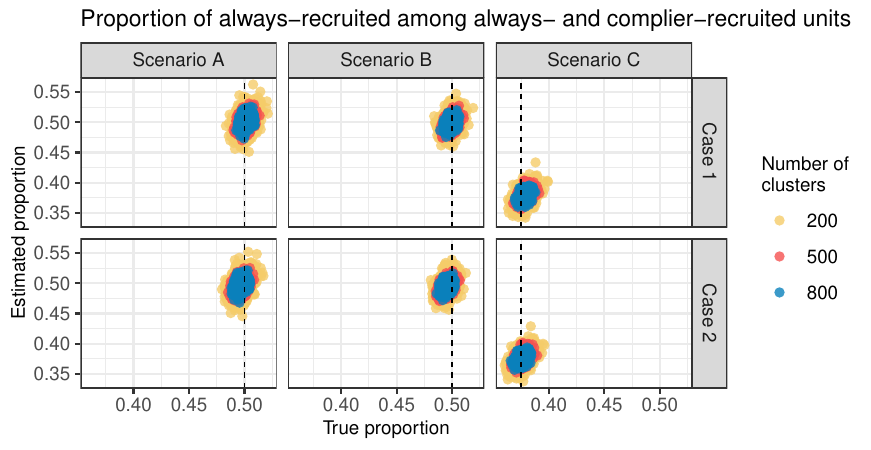}
\caption{True sample proportion of always-recruited among the always- and complier recruited individuals (x-axis) and estimated proportion (y-axis) across 500 data sets and $J \in \{200, 500, 800\}$ number of clusters (shown in color) organized across the scenario and cases in our simulations. The vertical lines correspond to approximate theoretical values for the true super-population proportion.}
\label{fig:sim_proportion_always}
\end{figure}

{The proportion of always-recruited among the always- and incentivized-recruited in \cref{eq:ident_prop_alw} is well-estimated across scenarios: \cref{fig:sim_proportion_always} shows the true and estimated values in our simulations, organized by scenario and case. The vertical lines show the theoretical values for the true proportion. We observe that there is variability around the true proportion from data set to data set, but the estimation procedure is accurate, and it improves as the number of clusters increases.}
Furthermore, the pseudo-likelihood estimators for the working propensity score parameters are unbiased with increasing precision under a larger number of clusters (as shown in 
Figure S.5
of the Supplement).
\cref{fig:bias_estPS_r1} shows the difference between the estimated and the true causal effect when using the proposed estimators based on the estimated working propensity score model across 500 simulated data sets and the various simulation scenarios under a balanced design. We notice that the causal estimators for the effect on the recruited population, and the always-recruited and incentivized-recruited subpopulations are unbiased throughout, and they become more precise when the number of clusters increases. This illustrates that our approach accurately estimates the causal effect for the recruited population, and two interpretable subsets of the overall population. The causal estimators based on the true propensity score, or under imbalanced designs are also unbiased; these results are shown in the supplement.

\begin{figure}
\centering
\includegraphics[width = \textwidth]{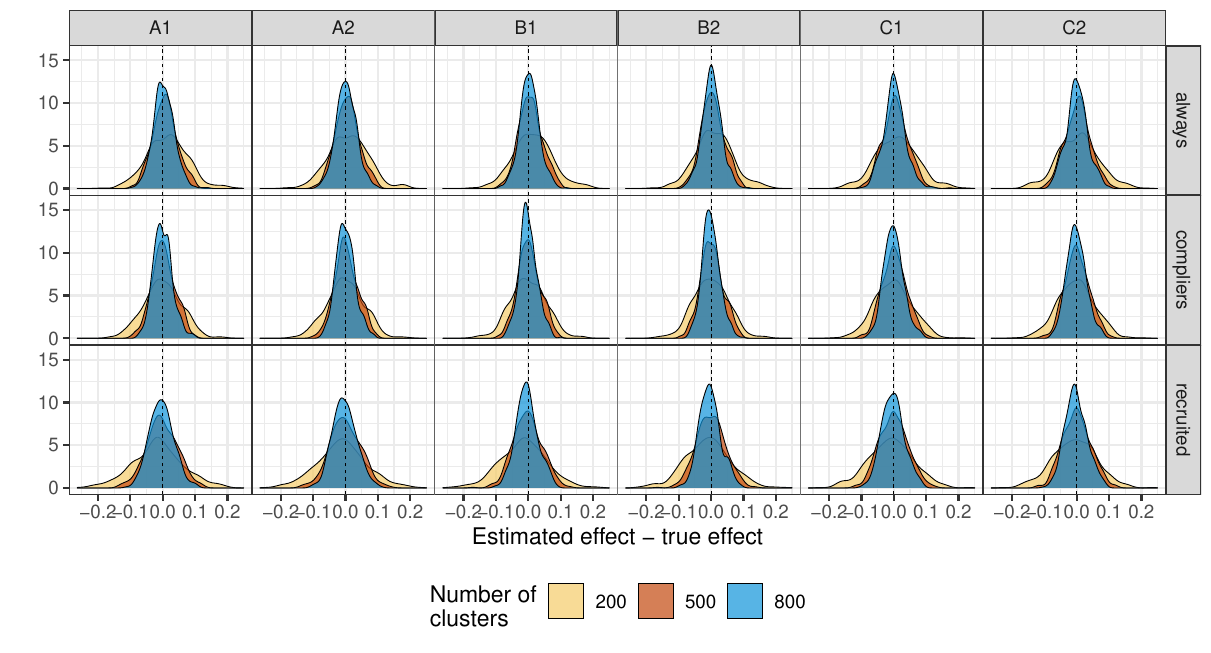}
\caption{True minus estimated causal effect among the always-recruited, the incentivized-recruited, and the recruited populations, across 500 data sets and under the 6 scenario-case combinations, and 3 choices for the number of clusters, when the probability of cluster treatment is 0.5.}
\label{fig:bias_estPS_r1}
\end{figure}

The coverage of the asymptotic 95\% intervals for the estimator that uses the known propensity score varied from 90 to 97\% across all scenarios and estimands. 
The coverage of the bootstrap 95\% intervals varied from 92.7 to 98.6\% for the estimands defined on the recruited population, and the always-recruited subpopulation. Coverage of 95\% intervals based on the bootstrap was lower but still reasonable for the average causal effect on the incentivized-recruited subpopulation, ranging from 85 to 93\%. A figure for the coverage across simulation scenarios is shown in the supplement.

\section{Application to the ARTEMIS Study} \label{sec:application}

Among the original 10,976 patients, we excluded hospitals whose sample size is less than 15, resulting in a final sample size of 10,400 from 203 hospitals with 108 intervention hospitals with 6,254 patients in total and 95 control hospitals with 4,146 patients in total. 
We compared the covariate distribution for hospital- and patient-level covariates among the recruited, and found that a number of important covariates, such as race, education and prior P2Y$_{12}$ use, were not balanced between arms. Details are given in Supplement 
E. 
This observation implies potential differential recruitment due to treatment assignment. Since the intervention hospitals offered vouchers for P2Y$_{12}$ to the enrolled patients without out-of-pocket cost and the control hospitals offered nothing, the monotonicity assumption is considered plausible.

We focused on the outcome of binary medication persistence (equal to 1 if the patient continued on the medication) as reported by the patient at their 1-year follow-up appointment. 
The difference of the mean outcome among the recruited treated and control patients is $2.75\%$ with 95\% confidence interval (CI), $0.59$ to $5.05\%$. We estimated the working propensity score model based on the pseudo-likelihood technique detailed in \cref{subsec:wps_estimation} and using all the available covariates listed in 
Table S.4
of the Supplement. Using the proposed method, we estimated the proportion (and 95\% confidence interval) of always-recruited among the always- and incentivized-recruited individuals to be $\widehat \nu = 0.754$ (95\% CI: $0.56$ to $1$).  We estimated the causal effect on the always-recruited to be $2.83\%$ (95\% CI: $0.87$ to $4.75\%$), on the incentivized-recruited to be $2.63\%$ (95\% CI: $-3.55$ to $4.41\%$), and on the recruited population to be $2.80\%$ (95\% CI: $0.6$ to $4.44\%$). The results imply potentially heterogeneous treatment effects across subpopulations. That is, the voucher appears to increase P2Y$_{12}$ medication persistence among the always-recruited patients with the associated confidence interval excluding zero. However, it may have had a weaker effect on the incentivized-recruited as the confidence interval is much wider and includes the null. Therefore, the effect in the recruited population is largely driven by the always-recruited subpopulation. 

From a policy standpoint, to achieve better medication persistence, resources should be allocated to the always-recruited, who are arguably more motivated to take the medication.
{Unfortunately, principal stratum membership for each individual is only partially observed, rendering exact targeting of the always-recruited individuals somewhat challenging. We study the covariate distribution of the always-recruited in comparison to the recruited population to obtain additional insights into this important subpopulation.}
Table \ref{tab:Xmean-strata} displays the weighted mean of each covariate in the always- and incentivized-recruited strata. Notable difference exists in some covariates between the two strata. Specifically, the always-recruited are more likely to be younger, women, non-White, have college education, have comorbidities, and not have had prior coronary artery bypass graft (CABG) surgery. {Therefore, implicit targeting of the always-recruited subpopulation can be performed by targeting patients with such covariate profiles.}
In addition, we found that these results may be sensitive to the violations of the ignorable recruitment assumption, with minimum $\Gamma$ value (for which the bounds include zero) equal to 1.13 for the always-recruited, and only 1.02 for the incentivized-recruited.

\begin{table}[hbtp]
    \centering
       \caption{Mean and standard deviation of the covariates among the recruited population, the always-recruited population and the incentivized-recruited subpopulation in the ARTEMIS.}
    \begin{tabular}{ccccc}
        \toprule
         Covariates & Recruited & Always Recruited & Incentivized Recruited \\
         \midrule
         Age & 61.74 (11.61) & 61.53 (11.59) & 62.35 (11.64) \\
         Male & 0.687 (0.464) & 0.645 (0.479) & 0.806 (0.396) \\
         White & 0.885 (0.319) & 0.862 (0.345) & 0.951 (0.217) \\
         Black & 0.096 (0.294) & 0.114 (0.318) & 0.044 (0.204) \\
         College & 0.510 (0.500) & 0.536 (0.499) & 0.437 (0.496) \\
         Prior MI & 0.201 (0.401) & 0.201 (0.401) & 0.202 (0.402) \\
         Private insurance & 0.641 (0.480) & 0.646 (0.478) & 0.627 (0.484) \\
         Prior P2Y$_{12}$ & 0.141 (0.348) & 0.143 (0.350) & 0.134 (0.341) \\
         Hemoglobin Level & 12.843 (2.046) & 12.875 (2.032) & 12.751 (2.082) \\
         Hypertension & 0.686 (0.464) & 0.716 (0.451) & 0.603 (0.489) \\
         DES use & 0.811 (0.391) & 0.810 (0.392) & 0.815 (0.389) \\
         Diabetes & 0.324 (0.468) & 0.333 (0.471) & 0.298 (0.458) \\
         Employment & 0.477 (0.500) & 0.483 (0.500) & 0.462 (0.499) \\
         Multivessel disease & 0.465 (0.499) & 0.460 (0.498) & 0.478 (0.500) \\
         Prior CABG surgery & 0.014 (0.118) & 0.012 (0.109) & 0.021 (0.142) \\
         \bottomrule
    \end{tabular}
 
    \label{tab:Xmean-strata}
\end{table}

{Furthermore, we estimated the effect on the always- and incentivized-recruited---equal to the effect on the treated recruited population---to be 2.78\% (95\% CI: 0.43 to 4.37\%). In ARTEMIS, individuals in the treated hospitals were easily recruited because the intervention allows them to acquire their medication without co-payment. Therefore, the recruited patient population in the treated hospitals is arguably more representative of the overall population, and the effect on the treated recruited subpopulation is likely to be well representative of the effect on the overall population. Therefore, our analysis suggests that removing co-payment barriers increases the persistence of P2Y$_{12}$ inhibitor for patients with myocardial infraction. This conclusion is based on the presumption that recruitment in treated clusters is almost universal, something that should be evaluated conceptually separately in each study.}

\section{Discussion} \label{sec:discussion}

Due to logistical and practical constraints, cluster randomized experiments are prone to post-randomization selection bias. In this paper, we established the conditions under which cluster randomization implies individual randomization, and clarified the different target populations and corresponding estimands. We provided weighting-based nonparametric identification results for these estimands based solely on the recruited sample. We also developed an interpretable sensitivity analysis method to assess the impact due to violation of the ignorable recruitment assumption. 

We have focused on addressing the identification challenges due to non-random recruitment, but assumed away other potential complications that might arise in cluster randomized experiments. Importantly, we have operated under the assumption of constant cluster membership such that the treatment assignment does not affect individual choice of clusters. This assumption effectively reduces the number of possible principal strata and is plausible in scenarios where the sampling frame for each cluster is determined before the start of the study and thus the cluster membership is a pre-treatment characteristic. \citet{schochet2022estimating,schochet2023estimating} discussed scenarios where the treatment assignment affects cluster membership, in which case the causal estimands are redefined and stronger identification assumptions are invoked for point identification. It would be desirable to expand our work to address treatment-dependent cluster membership and sensitivity methods for these stronger assumptions. 
{Finally, it remains an open question whether identifiability of overall effects can be achieved under weaker assumptions than the ones employed here. 
We found that non-differential recruitment and monotonicity (Assumptions \ref{ass:po_diffenrol} and \ref{ass:monotonicity}) 
imply that 
$\stratum[] 
\independent
\out[](0), \out[](1) \mid \covs[], \stratum[] \in \{a, c\}$ 
(see Proposition S.3), 
which is related to our identification \cref{ass:enrol_indep} but weaker, since it involves individuals in only two of the principal strata. Some preliminary simulation studies indicate that this weaker assumption might suffice for identification (see Supplement 
D),
and work by \cite{schochet2023estimating} involves an identifying ignorability assumption that is related to the one here.
Further theoretical work in this direction is warranted.
}

\section*{Acknowledgements}
This research was supported, in part, by the Patient-Centered Outcomes Research Institute (PCORI) contract ME-2019C1-16146. The contents of this article are solely the responsibility of the authors and do not necessarily represent the view of PCORI.

\clearpage

\allowdisplaybreaks
\setcounter{MaxMatrixCols}{12}

\begin{center}
\large
Supplementary materials for \\ ``\usetitle'' \\ by \\ \useauthorshort
\end{center}

\appendix
\renewcommand{\thetheorem}{S.\arabic{theorem}}
\renewcommand{\theproposition}{S.\arabic{proposition}}
\renewcommand{\theassumption}{S.\arabic{assumption}}
\renewcommand{\thelemma}{S.\arabic{lemma}}
\renewcommand{\theequation}{S.\arabic{equation}}
\renewcommand{\thetable}{S.\arabic{table}}
\renewcommand{\thefigure}{S.\arabic{figure}}
\setcounter{table}{0}
\setcounter{figure}{0}
\setcounter{equation}{0}
\setcounter{proposition}{0}
\setcounter{lemma}{0}
\setcounter{assumption}{0}
\setcounter{theorem}{0}

\section{Notation}

To facilitate clear presentation, we first list in \cref{supp_tab:notation} the notation that is used throughout the manuscript and supplement.

\begin{longtable}{R{0.25\textwidth}|L{0.64\textwidth}}
	\caption{Glossary of notation.} \\
	$J$ & Total number of clusters \\
	$\trtc, \covsc$ & Cluster level treatment and covariates \\
	$N_j, N$ & Number of individuals in the overall population for cluster $j$ and in all the clusters \\
	$n_j, n$ & Number of recruited individuals in cluster $j$ and in all the clusters \\
	$\trt, \covss, X_{ij}$ & Unit-level treatment, covariates, and the vector of both unit and cluster-level covariates \\
	$\enrol(0), \enrol(1), \enrol$ & Potential and observed values of the unit-level recruitment \\
	$\out(0), \out(1), \out$ & Potential and observed values of the outcome \\
	$\tau^O, \tau^R$ & Average effects on the overall and recruited populations \\
	$\tau^R_z$ & Average effect on the recruited populations with treatment level $z$ \\
	$\tau^O_s$ & Average effects for individuals in the overall population with recruitment principal stratum $S = s$ \\
	$\sum_{ij}$ & Represents a sum over the recruited individuals, for example $\sum_{ij} \out =  \sum_{j = 1}^J \sum_{i = 1}^{n_j} \enrol \out $
	\label{supp_tab:notation}
\end{longtable}

We use $i \in j$ to denote the individuals within cluster $j$ (recruited or not), and $i \in j^*$ to denote the recruited subjects in cluster $i$. Then, summations of the form $\sum_{i = 1}^N$ can be re-written as $\sum_{j = 1}^{J} \sum_{i \in j}$, and the summations of the form $\sum_{i = 1}^n$ can be re-written as $\sum_{j = 1}^J \sum_{i \in j^*}$.

\section{Proofs}
\label{supp:proofs}

\subsection{Randomized treatment on the overall population}

\begin{proof}[Proof of \cref{prop:overall}]
	Here, it is more convenient to use the single index notation. Specifically, let $i = 1, 2, \dots, N$ denote a randomly chosen unit in the overall population of the $J$ clusters, and let $\cluster = j$ denote that unit $i$ belongs to cluster $j$. 
	First, we start by showing that the cluster-level randomization of treatment, also implies that the cluster-level treatment is independent of cluster memberships (based on \cref{ass:choosing_cluster}), formalized as:
	\begin{equation}
	\Pr(\trtc[j] = 1 \mid \{\covsc\}_{j = 1}^J, \{ \covss[i], \out[i](0), \out[i](1) \}_{i = 1}^N, \cluster) = \Pr(\trtc[j] = 1).
	\label{eq:random_cluster_trt_withj}
	\end{equation}
	In other words, knowing which cluster subject $i$ belongs to does not provide any information on the cluster treatment level. 
	Note that \cref{ass:choosing_cluster} implies that, among the overall population,
	\[\Pr(Q_i = j \mid \{Z_j^c, \covsc\}_{j = 1}^{J}, \{ V_i, Y_i(0), Y_i(1)\}_{i = 1}^N ) =
	\Pr(Q_i = j \mid \{\covsc\}_{j = 1}^{J},  \{ V_i, Y_i(0), Y_i(1)\}_{i = 1}^N ).\] 
	Then
	\begin{align*}
	\Pr (\trtc[j] = 1 & \mid \{\covsc\}_{j = 1}^J, \{ \covss[i], \out[i](0), \out[i](1)\}_{i = 1}^N, \cluster) \\ 
	&= \frac{\Pr (\cluster \mid \trtc[j] = 1, \{\covsc\}_{j = 1}^J, \{\covss[i], \out[i](0), \out[i](1)\}_{i = 1}^N )}
	{\Pr (\{\covsc\}_{j = 1}^J, \{ \covss[i], \out[i](0), \out[i](1)\}_{i = 1}^N, \cluster)} \times \\
	& \hspace{30pt} \times
	\Pr (\trtc[j] = 1 \mid \{\covsc\}_{j = 1}^J, \{\covss[i], \out[i](0), \out[i](1)\}_{i = 1}^N ) \ 
	\Pr (\{\covsc\}_{j = 1}^J, \{\covss[i], \out[i](0), \out[i](1)\}_{i = 1}^N ) \\
	&= \frac{\Pr (\cluster \mid \{\covsc\}_{j = 1}^J, \{\covss[i], \out[i](0), \out[i](1)\}_{i = 1}^N )}
	{\Pr(\{\covsc\}_{j = 1}^J, \{\covss[i], \out[i](0), \out[i](1)\}_{i = 1}^N, \cluster)} \times 
	\tag{From \cref{ass:choosing_cluster}} \\
	& \hspace{30pt} \times
	\Pr (\trtc[j] = 1 \mid \{\covsc\}_{j = 1}^J, \{\covss[i], \out[i](0), \out[i](1)\}_{i = 1}^N ) \ 
	\Pr (\{\covsc\}_{j = 1}^J, \{ \covss[i], \out[i](0), \out[i](1)\}_{i = 1}^N ) \\
	&= \Pr (\trtc[j] = 1) \tag{From \cref{ass:randomization}}
	\end{align*}
	We have:
	\begin{align*}
	& \Pr (\trt[i] = 1  \mid \{\covsc\}_{j = 1}^J, \{ \covss[i], \out[i](0), \out[i](1)\}_{i = 1}^N ) = \\
	&= \sum_{j = 1}^J \Pr (\trt[i] = 1 \mid \{\covsc\}_{j = 1}^J, \{ \covss[i], \out[i](0), \out[i](1)\}_{i = 1}^N, \cluster = j) \Pr (\cluster = j \mid  \{\covsc\}_{j = 1}^J, \{ \covss[i], \out[i](0), \out[i](1)\}_{i = 1}^N ) \\
	&= \sum_{j = 1}^J \Pr (\trtc[j] = 1 \mid \{\covsc\}_{j = 1}^J, \{ \covss[i], \out[i](0), \out[i](1)\}_{i = 1}^N, \cluster = j) \Pr (\cluster = j \mid \{\covsc\}_{j = 1}^J, \{ \covss[i], \out[i](0), \out[i](1)\}_{i = 1}^N ) \\
	&= \Pr (\trtc[j] = 1) \sum_{j = 1}^J P(\cluster = j \mid \{\covsc\}_{j = 1}^J, \{ \covss[i], \out[i](0), \out[i](1)\}_{i = 1}^N ) \tag{From \cref{eq:random_cluster_trt_withj}} \\
	&= \Pr (\trtc[j] = 1)
	\end{align*}
	Further, \cref{eq:random_cluster_trt_withj} implies that $\Pr (\trtc[j] = 1 \mid \cluster) = \Pr (\trtc[j] = 1)$ and in turn
	\begin{align*}
	\Pr (\trt[i] = 1)
	&= \sum_{j = 1}^J P(\trt[i] = 1 \mid \cluster = j) \Pr(\cluster = j) \\
	&= \sum_{j = 1}^J  P(\trtc[j] = 1 \mid \cluster = j) \Pr (\cluster = j) \\
	&= \Pr (\trtc[j] = 1) \sum_{j = 1}^J \Pr (\cluster = j) \tag{From \cref{eq:random_cluster_trt_withj}}\\
	&= \Pr (\trtc[j] = 1).
	\end{align*}
	Putting these together: $\Pr(\trt[i] = 1 \mid \{\covsc\}_{j = 1}^J, \{ \covss[i], \out[i](0), \out[i](1)\}_{i = 1}^N ) = P(\trt[i] = 1) = P(\trtc[j] = 1)$.
	
\end{proof}

\subsection{ Relationships among assumptions for post-treatment recruitment}
\label{supp_subsec:assumptions}

\begin{proposition}
	The non-differential recruitment \cref{ass:po_diffenrol} is equivalent to subset ignorability, $\trt[] \independent \out[](0), \out[](1) \mid \covs[], \enrol[] = 1$.
	The ignorable recruitment \cref{ass:enrol_indep} is equivalent to conditional independence of principal stratum and potential outcomes $S \indep \{Y(0), Y(1)\} \mid X$.
	When \cref{ass:enrol_indep} holds, \cref{ass:po_diffenrol} also holds, though the reverse is not true.
	\label{prop:comparing_assumptions}
\end{proposition}

\begin{proof}
	
	First we show that \cref{ass:po_diffenrol} $\implies$ $\trt[] \independent \out[](0), \out[](1) \mid \covs[], \enrol[] = 1$. We denote $f_{\mathbf{\out[]}(\cdot)}$ as the distribution of $\mathbf{\out[]}(\cdot) = (\out[](0), \out[](1))$ and $\mathbf{y}(\cdot)$ as one realization.
	\begin{align*}
	f_{\mathbf{\out[]}(\cdot)} & (\mathbf{y}(\cdot) \mid \enrol[] = 1, \trt[] = \trtr, \covs[] = \covsr) \\
	&= \frac{P(\enrol[] = 1 \mid \trt[] = \trtr, \covs[] = \covsr, \mathbf{\out[]}(\cdot) = \mathbf{y}(\cdot)) P(\trt[] = \trtr \mid \covs[] = \covsr, \mathbf{\out[]}(\cdot) = \mathbf{y}(\cdot))  f_{\mathbf{\out[]}(\cdot)} (\mathbf{y}(\cdot) \mid \covs[] = \covsr) }
	{P(\enrol[]= 1 \mid \trt[] = \trtr, \covs[] = \covsr) P(\trt[] = \trtr \mid \covs[] = \covsr) }\\
	& = \frac{P(\enrol[] = 1 \mid \trt[] = \trtr, \covs[] = \covsr, \mathbf{\out[]}(\cdot) = \mathbf{y}(\cdot))}{P(\enrol[]= 1 \mid \trt[] = \trtr, \covs[] = \covsr)} f_{\mathbf{\out[]}(\cdot)} (\mathbf{y}(\cdot) \mid \covs[] = \covsr),
	\tag{Randomized treatment in the overall population -- See \cref{prop:overall}}
	\end{align*}
	From \cref{ass:po_diffenrol},
	\begin{align*}
	P(\enrol[] = 1 & \mid \trt[] = 1, \covs[] = \covsr) = \\
	&= \int P(\enrol[] = 1 \mid \trt[] = 1, \covs[] = \covsr, \mathbf{\out[]}(\cdot) = \mathbf{y}(\cdot)) f_{\mathbf{\out[]}(\cdot)}(\mathbf{y}(\cdot) \mid  \trt[] = 1, \covs[] = \covsr) \mathrm{d} \mathbf{y}(\cdot) \\
	&=  \delta(\covsr) \int P(\enrol[] = 1 \mid \trt[] = 0, \covs[] = \covsr, \mathbf{\out[]}(\cdot) = \mathbf{y}(\cdot)) f_{\mathbf{\out[]}(\cdot)}(\mathbf{y}(\cdot) \mid  \trt[] = 0, \covs[] = \covsr) \mathrm{d} \mathbf{y}(\cdot) \tag{Randomized treatment} \\
	&= \delta(\covsr) P(\enrol[] = 1 \mid \trt[] = 0, \covs[] = \covsr)
	\end{align*}
	Therefore,
	\begin{align*}
	f_{\mathbf{\out[]}(\cdot)}&(\mathbf{y}(\cdot) \mid \enrol[] = 1, \trt[] = 1, \covs[] = \covsr) =\\
	& =
	\frac{ \delta(\covsr) P(\enrol[] = 1 \mid \trt[] = 0, \covs[] = \covsr, \mathbf{\out[]}(\cdot) = \mathbf{y}(\cdot))}
	{\delta(\covsr) P(\enrol[] = 1 \mid \trt[] = 0, \covs[] = \covsr)}
	f_{\mathbf{\out[]}(\cdot)}(\mathbf{y}(\cdot) \mid \covs[] = \covsr)\\
	&= f_{\mathbf{\out[]}(\cdot)}(\mathbf{y}(\cdot) \mid \enrol[] = 1, \trt[] = 0, \covs[] = \covsr),
	\end{align*}
	which gives us that $\trt[] \independent \out[](0), \out[](1) \mid \covs[], \enrol[] = 1$. \\[20pt]
	Next we show that $\trt[] \independent \out[](0), \out[](1) \mid \covs[], \enrol[] = 1$ $\implies$ \cref{ass:po_diffenrol}. Consider
	\begin{align*}
	& \frac{P(\enrol[] = 1 \mid \covs[], \out[](0), \out[](1), \trt[] = 1)}
	{P(\enrol[] = 1 \mid \covs[], \out[](0), \out[](1), \trt[] = 0)} = \\
	&=
	\frac{P(\trt[] = 1 \mid \covs[], \out[](0), \out[](1), \enrol[] = 1)P(\enrol[] = 1 \mid \covs[], \out[](0), \out[](1)) \ \big/ \ P(\trt[] = 1 \mid \covs[], \out[](0), \out[](1))}
	{P(\trt[] = 0 \mid \covs[], \out[](0), \out[](1), \enrol[] = 1)P(\enrol[] = 1 \mid \covs[], \out[](0), \out[](1)) \ \big/ \ P(\trt[] = 0 \mid \covs[], \out[](0), \out[](1))} \\
	&=
	\frac{P(\trt[] = 1 \mid \covs[], \out[](0), \out[](1), \enrol[] = 1)}
	{P(\trt[] = 0 \mid \covs[], \out[](0), \out[](1), \enrol[] = 1)} 
	\frac{P(\trt[] = 0 \mid \covs[], \out[](0), \out[](1))}{P(\trt[] = 1 \mid \covs[], \out[](0), \out[](1))}
	\\
	&= \frac{P(\trt[] = 1 \mid \covs[], \enrol[] = 1)}
	{P(\trt[] = 0 \mid \covs[], \enrol[] = 1)}
	\frac{P(\trt[] = 0)}{P(\trt[] = 1)}
	,
	\end{align*}
	where the last equation holds from subset ignorability for the first ratio, and from \cref{prop:overall} for the second ratio. Therefore, the ratio of enrollment probabilities is only a function of the covariates and \cref{ass:po_diffenrol} is satisfied. \\[20pt]
	Next we show that \cref{ass:enrol_indep} $\implies S \indep \{Y(0), Y(1)\} \mid Z, X$. 
	Notice that $\mathbb{I}(S \in \{a, c\}) = ZR$, $\mathbb{I}(S \in \{a, d\}) = (1 - Z)R$. Therefore, $R \indep \{Y(0), Y(1)\} \mid Z, X$ implies $ (\mathbb{I}(S \in \{a, c\}), \mathbb{I}(S \in \{a, d\}))\indep \{Y(0), Y(1)\} \mid Z, X$. 
	The entries of $(\mathbb{I}(S \in \{a, c\}), \mathbb{I}(S \in \{a, d\}))$ are a 1-1 map with the stratum membership $S$ since
	\[
	(\mathbb{I}(S \in \{a, c\}), \mathbb{I}(S \in \{a, d\})) = \begin{cases}
	(0, 0) & \iff S = n \\
	(1, 0) & \iff S = c \\
	(0, 1) & \iff S = d \\
	(1, 1) & \iff S = a.
	\end{cases}
	\]
	Therefore, we have that $S \indep \{Y(0), Y(1)\} \mid Z, X$.
	Both $S$ and $\{Y(0), Y(1)\}$ are independent of $Z$, and hence we may drop $Z$ from the conditioning set. \\[20pt]
	Since all the steps also work backwards, we have that $S \indep \{Y(0), Y(1)\} \mid Z, X \implies$ \cref{ass:enrol_indep}, which completes the proof that the two assumptions are equivalent. \\[20pt]
	Next we show that \cref{ass:enrol_indep} implies \cref{ass:po_diffenrol}. From \cref{ass:enrol_indep} we have that
	\[
	\frac{P(\enrol[] = 1 \mid \covs[], \out[](0), \out[](1), \trt[] = 1)}
	{P(\enrol[] = 1 \mid \covs[], \out[](0), \out[](1), \trt[] = 0)}
	= 
	\frac{P(\enrol[] = 1 \mid \covs[], \trt[] = 1)}
	{P(\enrol[] = 1 \mid \covs[], \trt[] = 0)}
	\]
	is only a function of covariates and \cref{ass:po_diffenrol} is satisfied. \\[20pt]
	The reverse is not true, and \cref{ass:po_diffenrol} does not imply \cref{ass:enrol_indep}. The following example is a situation where \cref{ass:po_diffenrol} holds, and \cref{ass:enrol_indep} does not. For a binary outcome, assume that
	\[
	P(\enrol[] = 1 \mid \covs[], \out[](0), \out[](1), \trt[]) =
	\begin{cases}
	0.5 & \text{if } \out[](0) = 0 \text{ and } \trt[] = 1, \\
	0.2 & \text{if } \out[](0) = 0 \text{ and } \trt[] = 0, \\
	0.75 & \text{if } \out[](0) = 1 \text{ and } \trt[] = 1, \\
	0.3 & \text{if } \out[](0) = 1 \text{ and } \trt[] = 1.
	\end{cases}
	\]
\end{proof}

\subsection{Additional information on the non-differential recruitment functional}
\label{supp_subsec:delta_function}

In \cref{prop:implication_diffenrol}, we establish equivalent forms of the ratio of recruitment probabilities defined in equation \cref{eq:delta_def} of the manuscript.

\begin{proposition}
	Under constant cluster membership (\cref{ass:choosing_cluster}) and treatment randomization at the cluster level (\cref{ass:randomization}), the ratio of recruitment probabilities in \cref{eq:delta_def} is equal to
	\begin{align*}
	&\frac{\Pr(\enrol = 1 \mid \out(0) = \outr_0, \out(1) = \outr_1, X_{ij}=x, \trt = 1)}{\Pr(\enrol = 1 \mid \out(0) = \outr_0, \out(1) = \outr_1, X_{ij}=x, \trt = 0)} = \\
	& \hspace{80pt}
	\frac{\Pr(S \in \{a, c\} \mid Y(1) = y_1, Y(0) = y_0, X = x)}{\Pr(S \in \{a, d\} \mid Y(1) = y_1, Y(0) = y_0, X = x)},
	\end{align*}
	for all values of $x, y_0, y_1$. Furthermore, if monotonicy of the recruitment indicator (\cref{ass:monotonicity}) holds, then this ratio is equal to
	\(
	1 / \Pr(S = a \mid S \in \{a, c\}, Y(1) = y_1, Y(0) = y_0, X = x) \geq 1.
	\)
	\label{prop:implication_diffenrol}
\end{proposition}

\begin{proof}
	Note that 
	\begin{align*}
	&\Pr(R = 1\mid Y(1), Y(0), X, Z = 1) \\
	=& \sum_{s \in \{a, c, d, n\}}\Pr(R = 1\mid Y(1), Y(0), X, Z = 1, S = s)\Pr(S = s \mid Y(1), Y(0), X, Z = 1) \\
	=& \sum_{s \in \{a, c\}}\Pr(S = s \mid Y(1), Y(0), X) \tag{From \cref{prop:overall}, we can drop $Z$ from the conditioning set.} \\
	=& \Pr(S \in \{a, c\}\mid Y(1), Y(0), X).
	\end{align*}
	Similarly, 
	\begin{equation*}
	\Pr(R = 1\mid Y(1), Y(0), X, Z = 0) = \Pr(S \in \{a, d\}\mid Y(1), Y(0), X).
	\end{equation*}
	Under \cref{ass:monotonicity}, the $d$ stratum does not exist, and $$\Pr(R = 1\mid Y(1), Y(0), X, Z = 0) = \Pr(S = a\mid Y(1), Y(0), X).$$
	The proof is immediate since
	$$\Pr(S = a \mid S \in \{a, c\}, Y(1), Y(0), X) = \frac{\Pr(S = a \mid Y(1), Y(0), X)}{\Pr(S \in \{a, c\} \mid  Y(1), Y(0), X)},$$
	and since any probability is between 0 and 1, we have that $1 / \Pr(S = a \mid S \in \{a, c\}, Y(1) = y_1, Y(0) = y_0, X = x) \geq 1.$
\end{proof}

\begin{proposition}
	Under constant cluster membership (\cref{ass:choosing_cluster}) and treatment randomization at the cluster level (\cref{ass:randomization}), if non-differential recruitment (\cref{ass:po_diffenrol}) and monotonicity (\cref{ass:monotonicity}) hold, then it holds that $\stratum \indep \out(0), \out(1) \mid \covs, \stratum \in \{a, c\}$.
	\label{prop:comparing_assumptions2}
\end{proposition}
\begin{proof}
	Since 
	$$
	\delta(x) = \frac{\Pr(\enrol = 1 \mid \out(0) = \outr_0, \out(1) = \outr_1, X_{ij}=x, \trt = 1)}{\Pr(\enrol = 1 \mid \out(0) = \outr_0, \out(1) = \outr_1, X_{ij}=x, \trt = 0)},
	$$
	we know from \cref{prop:implication_diffenrol}, and using monotinicity, that
	$$
	\delta(x) = 1 / \Pr(\stratum = a \mid \stratum \in \{a, c\}, \out(1) = y_1, \out(0) = y_0, \covs = x).
	$$
	Then, we have that
	\begin{align*}
	P(\stratum = a & \mid \covs, \stratum \in \{a, c\}) = \\
	&= \int_{y_0, y_1} P(\stratum = a \mid \covs, \out(0) = y_0, \out(1) = y_1, \stratum \in \{a, c\}) f(y_0, y_1 \mid \covs, \stratum \in \{a, c\}) \\
	&= \int_{y_0, y_1} \delta(x)^{-1} f(y_0, y_1 \mid \covs, \stratum \in \{a, c\}) \\
	&= \delta(x)^{-1},
	\end{align*}
	assuming for simplicity that the potential outcome distribution admits a density.
	So we have that
	\[
	\Pr(\stratum = a \mid \stratum \in \{a, c\}, \out(1) = y_1, \out(0) = y_0, \covs = x) =
	\Pr(\stratum = a \mid \stratum \in \{a, c\}, \covs = x).
	\]
	Since $\stratum$ is binary within the group $\stratum \in \{a, c\}$, this implies that \(\stratum \indep \out(0), \out(1) \mid \covs, \stratum \in \{a, c\}.\)
\end{proof}

\vspace{20pt}
The function $\delta(x)$ in \cref{ass:po_diffenrol} is defined as
\[
\delta(x) = \frac{\Pr(\enrol = 1 \mid \out(0) = \outr_0, \out(1) = \outr_1, X_{ij}=x, \trt = 1)}{\Pr(\enrol = 1 \mid \out(0) = \outr_0, \out(1) = \outr_1, X_{ij}=x, \trt = 0)},
\]
and its form under assumptions with respect to principal stratum membership was given in \cref{prop:implication_diffenrol}. We note here that the definition of $\delta(x)$ includes the values of the potential outcomes, and $\delta(x)$ need {\it not} necessarily be equal to the ratio of
\[
\frac{\Pr(\enrol = 1 \mid X_{ij}=x, \trt = 1)}{\Pr(\enrol = 1 \mid X_{ij}=x, \trt = 0)},
\]
or functions similar to those in \cref{prop:implication_diffenrol} excluding potential outcomes. The following proposition establishes that under {\it additional} assumptions, this equivalence holds.

\begin{proposition}
	Assume that the conditions of \cref{prop:implication_diffenrol} hold. Then,
	\begin{enumerate}[label=(\alph*)]
		\item If \cref{ass:enrol_indep} holds, then 
		\(
		\delta(x) = {\Pr(\stratum \in \{a, c\} \mid \covs = x)} / {\Pr(\stratum \in \{a, d\} \mid \covs = x)}.
		\)
		\item If \cref{ass:monotonicity} holds, then
		\(
		\delta(x) = 1 / P(\stratum = a \mid \stratum \in \{a, c\}, \covs = \covsr).
		\)
	\end{enumerate}
	Under either assumption,
	\( \displaystyle
	\delta(x) = \frac{\Pr(\enrol = 1 \mid X_{ij}=x, \trt = 1)}{\Pr(\enrol = 1 \mid X_{ij}=x, \trt = 0)}.
	\)
	\label{prop:delta_under_assumptions}
\end{proposition}
\begin{proof} $\ $
	\begin{enumerate}[label=(\alph*)]
		\item 
		\begin{align*}
		\delta(x) &= \frac{\Pr(\enrol = 1 \mid \out(0) = \outr_0, \out(1) = \outr_1, X_{ij}=x, \trt = 1)}{\Pr(\enrol = 1 \mid \out(0) = \outr_0, \out(1) = \outr_1, X_{ij}=x, \trt = 0)} \\
		&= \frac{\Pr(\enrol = 1 \mid X_{ij}=x, \trt = 1)}{\Pr(\enrol = 1 \mid X_{ij}=x, \trt = 0)} \tag{Using \cref{ass:enrol_indep} for the numerator and denominator} \\
		&= \frac{P(\stratum \in \{a, c\} \mid \covs = \covsr, \trt = 1)}
		{P(\stratum \in \{a, d\} \mid \covs = \covsr, \trt = 0)} \\
		&= \frac{P(\stratum \in \{a, c\} \mid \covs = \covsr)}
		{P(\stratum \in \{a, d\} \mid \covs = \covsr)} \tag{From \cref{prop:overall}}
		\end{align*}
		(Note that the second equation above is one of the results we are showing.)
		
		\item
		\begin{align*}
		P(\stratum & = a \mid \stratum \in \{a, c\}, \covs = \covsr) = \\
		&= 
		\int_{y_0, y_1} P(\stratum = a \mid \stratum \in \{a, c\}, \covs = \covsr) f(y_0, y_1 \mid \stratum \in \{a, c\}, \covs = \covsr) \mathrm{d}(y_0, y_1) \\
		&= \delta(\covsr)^{-1} \int_{y_0, y_1} f(y_0, y_1 \mid \stratum \in \{a, c\}, \covs = \covsr) \mathrm{d}(y_0, y_1) \tag{From \cref{prop:implication_diffenrol}} \\
		&= \delta(\covsr)^{-1}
		\end{align*}
		Therefore, we can also write
		\begin{align*}
		\delta(x) &= \frac1{P(\stratum = a \mid \stratum \in \{a, c\}, \covs = \covsr)} \\
		&= \frac{P(\stratum \in \{a, c\} \mid \covs = \covsr)}{P(\stratum = a \mid \covs = \covsr)} \\
		&= \frac{P(\stratum \in \{a, c\} \mid \covs = \covsr, \trt = 1)}{P(\stratum = a \mid \covs = \covsr, \trt = 0)} \tag{From \cref{prop:overall}} \\
		&= \frac{P(\enrol = 1 \mid \covs = \covsr, \trt = 1)}{P(\enrol = 1 \mid \covs = \covsr, \trt = 0)} \tag{Using monotonicity}
		\end{align*}
	\end{enumerate}
\end{proof}

The results in \cref{prop:implication_diffenrol} and \cref{prop:delta_under_assumptions} imply that under standard conditions (constant cluster membership and randomization of the treatment assignment) and monotonicity, the function $\delta(x)$ satisfies that
\[
\delta(\covsr) = \frac{P(\enrol = 1 \mid \covs = \covsr, \trt = 1)}{P(\enrol = 1 \mid \covs = \covsr, \trt = 0)} \geq 1.
\]

\subsection{The working propensity score}
\label{app_subsec:wps}

We show that enrollment positivity leads to working propensity score positivity.

\begin{proposition}[Working propensity score positivity]
	If \cref{ass:positivity_enrollment} holds, then there exists $\delta' > 0$ such that $\delta' < \wps < 1 - \delta'$ for all $\covsr$.
	\label{app_prop:positivity}
\end{proposition}
\begin{proof}
	Remember that
	\begin{align}
	\wps &= \frac{P(\enrol[] = 1 \mid \trt[] = 1, \covs[] = \covsr)}{\prob(\enrol[] = 1 \mid \covs[] = \covsr)} \prob(\trt[] = 1).
	\label{supp_eq:ps}
	\end{align}
	From \cref{ass:positivity_enrollment} we have that $P(\enrol[] = 1 \mid \trt[] = \trtr, \covs[] = \covsr) > \delta$, which gives us that
	\( \displaystyle \wps > \delta P(\trt[] = 1). \)
	It also implies that $P(\enrol[] = 1 \mid \covs[] = \covsr) > \delta$, which in turn gives us that
	\( \displaystyle \wps < P(\trt[] = 1) / \delta. \)
	By setting $\delta' = \min\{ \delta P(\trt[] = 1), 1 - P(\trt[] = 1) / \delta \}$, we have the result.
\end{proof}

Next, we show the relationship between the working propensity score $\wps$ and the ratio of recruitment probabilities $\delta(x)$ in \cref{ass:po_diffenrol}.

\begin{proposition}
	If the assumptions of constant cluster membership (\cref{ass:choosing_cluster}), treatment randomization (\cref{ass:randomization}), and non-differential recruitment (\cref{ass:po_diffenrol}) hold, then $\wps = \delta(x) / \{ \delta(x) + r^{-1} \}$ where $r = \Pr(\trtc[] = 1) / \Pr(\trtc[] = 0)$.
\end{proposition}
\begin{proof}
	In \cref{prop:comparing_assumptions}, it was established that \cref{ass:po_diffenrol} is equivalent to $\trt \indep \{\out(0), \out(1)\} \mid \covs, \enrol = 1$. Therefore,
	\begin{align*}
	\wps &= P(\trt[] = 1 \mid \covs[] = \covsr, \enrol[] = 1) \\
	&= P(\trt[] = 1 \mid \covs[] = \covsr, \enrol[] = 1, \out[](0), \out[](1)) \\
	&= \frac{P(\enrol[] = 1 \mid \covs[] = \covsr, \out[](0), \out[](1), \trt[] = 1)
		P(\trt[] = 1 \mid  \covs[] = \covsr, \out[](0), \out[](1))}
	{P(\enrol[] = 1 \mid \covs[] = \covsr, \out[](0), \out[](1))} \\
	&=
	\frac{1}{
		1 + \dfrac{P(\enrol[] = 1 \mid \covs[] = \covsr, \out[](0), \out[](1), \trt[] = 0)
			P(\trt[] = 0 \mid  \covs[] = \covsr, \out[](0), \out[](1))}{P(\enrol[] = 1 \mid \covs[] = \covsr, \out[](0), \out[](1), \trt[] = 1)
			P(\trt[] = 1 \mid  \covs[] = \covsr, \out[](0), \out[](1))}
	} \\
	&= \frac1{1 + \delta^{-1}(x) r^{-1}} \tag{Using \cref{prop:overall}}\\
	&= \frac{\delta(x)}{\delta(x) + r^{-1}}
	\end{align*}
\end{proof}


\subsection{Identifiability of average causal effects for the recruited population}
\label{supp_subsec:ident_recruited}

Here, we use the equivalence between weighting and conditioning estimands for our proof. The weighted average causal effect among a subpopulation of the recruited population with covariate distribution $g^R(x)$ is defined as
\[
\tau^R_g = \E[ h^R(X) [Y(1) - Y(0)] \mid \enrol[] = 1 ],
\]
where $h^R(x) = g^R(x) / f^R(x)$, and $f^R(x)$ is the covariate distribution among the recruited individuals.
For $g^R(x) = f^R(x)$, we have that $\tau_g^R = \tau^R$. Also, for $g^R(x) = f(x \mid \trt[] = 1, \enrol[] = 1)$ or $g^R(x) = f(x \mid \trt[] = 0, \enrol[] = 1)$, the estimand $\tau_g^R$ reverts to the average causal effect among the treated recruited individuals $\tau^R_1$ or the control recruited individuals $\tau^R_0$, respectively.

\begin{proposition}
	If \cref{ass:choosing_cluster}, \cref{ass:randomization}, \cref{ass:po_diffenrol}, and \cref{ass:positivity_enrollment} hold, then
	\begin{align*}
	\tau^R_g &= \E \left[ h^R(X) \left\{ \frac{\trt[] \out[]}{e(X)} - \frac{(1 - \trt[])\out[]}{1 - e(X)} \right\}\bigg | \enrol[] = 1 \right],
	\end{align*}
	for any subpopulation of the recruited population with covariate distribution $g^R(x)$ and corresponding tilting function $h^R(x)$.
	\label{prop:identifiability_recruited}
\end{proposition}

\begin{proof}
	%
	%
	%
	We have
	\begin{align*}
	\E \left\{ h^R(\covs[]) \frac{\trt[]\out[]}{\wpS} \mid \enrol[] = 1 \right\} &=
	\E \left\{ \E \left[ h^R(\covs[]) \frac{\trt[]\out[](1)}{\wpS} \mid \covs[], \out[](1), \out[](0), \enrol[] = 1 \right] \mid \enrol[] = 1 \right\} \\
	&=
	\E \left\{ \frac{h^R(\covs[])}{\wpS} \out[](1) \ \E \left[ \trt[] \mid \covs[], \out[](1), \out[](0), \enrol[] = 1 \right] \mid \enrol[] = 1 \right\} \\
	&= \E \left\{ \frac{h^R(\covs[])}{\wpS} \out[](1) \ P( \trt[] = 1 \mid \covs[], \out[](1), \out[](0), \enrol[] = 1 ) \mid \enrol[] = 1 \right\} \\
	&= \E \left\{ \frac{h^R(\covs[])}{\wpS} \out[](1) \ P( \trt[] = 1 \mid \covs[], \enrol[] = 1 ) \mid \enrol[] = 1 \right\}
	\tag{From \cref{prop:comparing_assumptions} \& using \cref{ass:po_diffenrol}} \\
	&= \E \left\{ h^R(\covs[]) \ \out[](1) \mid \enrol[] = 1 \right\}
	\end{align*}
	and similarly we can get that
	\[
	\E \left\{ h^R(\covs[]) \frac{(1 - \trt[])\out[]}{1 - \wpS} \mid \enrol[] = 1 \right\}
	= \E \left\{ h^R(\covs[]) \ \out[](0) \mid \enrol[] = 1 \right\},
	\]
	which verifies the weighting-based identification formula.
	
\end{proof}

\subsection{Identifiability of average causal effects for the overall population}
\label{supp_subsec:ident_overall}

We similarly use weighting definitions of estimands for the overall population for our proofs. The weighted average causal effect among a subpopulation of the overall population with covariate distribution $g^O(x)$ is defined as
\[
\tau^O_g = \E \left[ h^O(X) [Y(1) - Y(0)]  \right],
\]
where $h^O(x) = g^O(x) / f_X(x)$, and $f_X(x)$ is the covariate distribution in the overall population.
For $g^O(x) = f_X(x)$, we have that $\tau_g^O = \tau^O$. Also, for $g^O(x) = f(x \mid \stratum[] \in \{a, d\})$, the estimand $\tau_g^O$ reverts to the average causal effect among the always- and disincentivized-recruited individuals $\tau^O_{a,d}$, and similarly for the other principal strata. 

\begin{proof}[Proof of \cref{prop:identifiability_overall}]
	
	First, we establish how probabilistic statements about the recruitment indicator $\enrol[]$ given treatment can be linked to probabilistic statements about the potential recruitment value $\enrol[](z)$:
	\begin{equation}
	\begin{aligned}
	\prob(\enrol[] = 1 \mid \trt[] = \trtr, \covs[] = \covsr)
	&=
	\prob(\enrol[](\trtr) = 1 \mid \trt[] = \trtr, \covs[] = \covsr)  \\
	&= \frac{\prob(\trt[] = \trtr \mid \enrol[](z) = 1, \covs[] = \covsr)}{\prob(\trt[] = \trtr \mid \covs[] = \covsr)} \prob(\enrol[](z) = 1 \mid \covs[] = \covsr) \\
	&= \prob(\enrol[](z) = 1 \mid \covs[] = \covsr),
	\end{aligned}
	\label{app_eq:enrol_given_trt}
	\end{equation}
	using consistency of the potential recruitment indicators and that the treatment is randomized from \cref{prop:overall}. Similarly
	$$
	\prob(\enrol[] = 1 \mid \trt[] = \trtr) = \prob(\enrol[](z) = 1).
	$$
	For the following, we will also use the result from \cref{prop:comparing_assumptions} which shows that \cref{ass:enrol_indep} implies subset ignorability, $\trt[] \independent \out[](0), \out[](1) \mid \covs[], \enrol[] = 1$.
	
	\begin{enumerate}[label=(\alph*)]
		
		\item In this part, the subpopulation of the recruited individuals that is of interest corresponds to that with covariate density $g^R$ of the control individuals among the recruited population, $g^R(x) = f(x \mid \trt[] = 0, \enrol[] = 1)$. The corresponding tilting function on the recruited population is denoted as $h^R(x)$. This implies that $\tau_g^R = \tau^R_0$. Also, the covariate density of interest on the overall population, $g^O$, corresponds to the covariate density of the always and defying-recruiters, $g^O(x) = f(x \mid \stratum[] \in \{a, d\})$, with corresponding overall tilting function $h^O(x)$. Then $\tau^O_g = \tau^O_{a,d}$. First, we show that these two covariate distributions are the same.
		\begin{align*}
		g^R(\covsr) &= f_{\covs[]} (\covsr \mid \trt[] = 0, \enrol[] = 1) \\
		&= \frac{P(\enrol[] = 1 \mid \covs[] = \covsr, \trt[] = 0) P(\trt[] = 0 \mid \covs[] = \covsr)}{P(\enrol[] = 1 \mid \trt[] = 0) P(\trt[] = 0)} f_{\covs[]} (\covsr) \\
		&= \frac{P(\enrol[](0) = 1 \mid \covs[] = \covsr)}{P(\enrol[](0) = 1)} f_{\covs[]} (\covsr)
		\tag{From equation \cref{app_eq:enrol_given_trt} and \cref{prop:overall}} \\
		&= f_{\covs[]}(\covsr \mid \enrol[](0) = 1) \\
		&= f_{\covs[]}(\covsr \mid S \in \{a, d\}) \numberthis \label{app_eq:dist_R0_1} \\
		&= g^O(\covsr)
		\end{align*}
		
		Then, we show that
		$\E[h^R(\covs[]) \out[](\trtr) \mid \enrol[] = 1] = 
		\E[h^O(\covs[]) \out[](\trtr)]$,
		which implies that the estimands of these two populations are equal, in that
		\[
		\tau_g^R = \E[h^R(\covs[]) (\out[](1) - \out[](0)) \mid \enrol[] = 1]
		=
		\E[h^0(\covs[]) (\out[](1) - \out[](0))] = \tau_g^O
		\]
		To show this, consider the expected outcome in the tilted recruited population. Remember that this tilted population with tilting function $h^R$ represents the recruited controls:
		\begin{align*}
		\E [ h^R(\covs[]) \ \out[](1) \mid \enrol[] = 1 ] &=
		\int_{\covsr} \int_{\outr} \outr \ f(\out[](1) = \outr \mid \covs[] = \covsr[], \enrol[] = 1) \mathrm{d}\outr \ h^R(\covsr)  f(\covsr \mid \enrol[] = 1) \mathrm{d}\covsr \\
		&=
		\int_{\covsr} \int_{\outr} \outr \ { f(\out[](1) = \outr \mid \covs[] = \covsr[], \enrol[] = 1)} \mathrm{d}\outr\ g^R(\covsr) \ \mathrm{d}\covsr \\
		&=
		\int_{\covsr} \int_{\outr} \outr \ { f(\out[](1) = \outr \mid \trt[] = 1, \covs[] = \covsr[], \enrol[] = 1)} \mathrm{d}\outr \ g^O(\covsr) \mathrm{d}\covsr
		\tag{From subset ignorability and the result above} \\
		&=
		\int_{\covsr} \int_{\outr} \outr \ { f(\out[](1) = \outr \mid \trt[] = 1, \covs[] = \covsr[])} \mathrm{d}\outr \ h^O(\covsr)  f^O_{\covs[]}(\covsr) \mathrm{d}\covsr
		\tag{From \cref{ass:enrol_indep}} \\
		&=
		\int_{\covsr} \int_{\outr} \outr \ { f(\out[](1) = \outr \mid \covs[] = \covsr[])} \mathrm{d}\outr \ h^O(\covsr)  f^O_{\covs[]}(\covsr) \mathrm{d}\covsr
		\tag{From \cref{prop:overall}} \\
		&= \E[ h^O(\covs[]) \out[](1)].
		\end{align*}
		We can similarly show that 
		$\E [ h^R(\covs[]) \ \out[](0) \mid \enrol[] = 1 ] =
		\E [ h^O(\covs[]) \ \out[](0) ].
		$
		Therefore, $\tau^R_0 = \tau^R_g = \tau^O_g = \tau^O_{a,d}$.
		
		\item As long as we show that $g^R(\covsr) = g^O(\covsr)$ for $g^R(\covsr) = f(\covsr \mid \trt[] = 1, \enrol[] = 1)$ and $g^O(\covsr) = f(\covsr \mid \stratum[] \in \{a, c\})$, then the proof will follow similarly to the steps for the first part. Here
		\begin{align*}
		g^R(\covsr) &= f_{\covs[]} (\covsr \mid \trt[] = 1, \enrol[] = 1) \\
		&= \frac{P(\enrol[] = 1 \mid \covs[] = \covsr, \trt[] = 1) P(\trt[] = 1 \mid \covs[] = \covsr)}{P(\enrol[] = 1 \mid \trt[] = 1) P(\trt[] = 1)} f_{\covs[]} (\covsr) \\
		&= \frac{P(\enrol[](1) = 1 \mid \covs[] = \covsr)}{P(\enrol[](1) = 1)} f_{\covs[]} (\covsr)
		\tag{From equation \cref{app_eq:enrol_given_trt} and \cref{prop:overall}} \\
		&= f_{\covs[]}(\covsr \mid \enrol[](1) = 1) \\
		&= f_{\covs[]}(\covsr \mid S \in \{a, c\}) \\
		&= g^O(\covsr)
		\end{align*}
		
	\end{enumerate}
	
	The identification formulas for $\tau^O_{a,d}$ and $\tau^O_{a,c}$ are straightforward to acquire based on the formulas for the formulas for $\tau^R_0$ and $\tau^R_1$ in \cref{prop:identifiability_recruited}.
\end{proof}

\begin{proof}[Proof of \cref{theorem:compliers}]
	\begin{align*}
	p^t &= P(\trt[] = 1 \mid \enrol[] = 1) \\
	&= \frac{P(\enrol[] = 1 \mid \trt[] = 1) P(\trt[] = 1)}{P(\enrol[] = 1 \mid \trt[] = 1) P(\trt[] = 1) + P(\enrol[] = 1 \mid \trt[] = 0) P(\trt[] = 0)} \\
	&= \frac{P(\stratum[] \in \{a, c\} \mid \trt[] = 1) \pi^t}{P(\stratum[] \in \{a, c\} \mid \trt[] = 1) \pi^t + P(\stratum[] = a \mid \trt[] = 0) (1 - \pi^t)} \\
	&= \frac{(\pi_a + \pi_c) \pi^t}{(\pi_a + \pi_c) \pi^t + \pi_a (1 - \pi^t)},
	\end{align*}
	where the second equation using monotonicity and \cref{prop:overall}, and the third equation uses the definitions of $\pi_{a,c}, \pi_a$ and \cref{prop:overall} that states that the treatment is independent of the stratum memberships. Solving for $\pi_a / (\pi_a + \pi_c)$ gives us the result. Based on this and \cref{prop:identifiability_overall}, $\tau^O_c$ is also identifiable.
\end{proof}

\subsection{Asymptotic normality of causal effect estimators}
\label{supp_subsec:asym_norm}

We establish the value of $1 - \wps$ which will be useful moving forward. Note that
\begin{equation}
\begin{aligned}
\wps &= \frac{P(\enrol[] = 1 \mid \trt[] = 1, \covs[] = \covsr)}{\prob(\enrol[] = 1 \mid \covs[] = \covsr)} \prob(\trt[] = 1) \\
\iff 1 - \wps
&= \frac{\prob(\enrol[] = 1 \mid \covs[] = \covsr) - P(\enrol[] = 1 \mid \trt[] = 1, \covs[] = \covsr) \prob(\trt[] = 1 \mid \covs[] = \covsr)}{\prob(\enrol[] = 1 \mid \covs[] = \covsr)} \\
&= 
\frac{P(\enrol[] = 1 \mid \trt[] = 0, \covs[] = \covsr)}{\prob(\enrol[] = 1 \mid \covs[] = \covsr)} \prob(\trt[] = 0)
\end{aligned}
\label{app_eq:1_wps}
\end{equation}

\begin{assumption}[Regularity assumptions for asymptotic normality] $\ $
	\begin{enumerate}[label=(\alph*)]
		
		
		\item $N_j$ is independent of the cluster average potential outcome, i.e., if the cluster average potential outcome is defined as
		\( \overline Y_{j,h}^O = \frac1{N_j} \sum_{i \in j}  \out(1)  h^O(\covs) \) for some tilting function $h^O(x)$, then we have that
		\( N_j \independent \overline Y_{j,h}^O \),
		\label{app_ass_item:n_po_indep}
		
		\item there exists $M_N$ such that $P(N_j < M_N) = 1$, \label{app_ass_item:bounded_n}
		
		
		\item there exists $M_Y$ such that $P(|\out| < M_Y) = 1$, \label{app_ass_item:bounded_Y}
		
	\end{enumerate}
	
	
	\label{app_ass:regularity}
\end{assumption}

\begin{proof}[Proof of \cref{prop:asym_norm}]
	
	To show asymptotic normality of our estimator we will use Theorem 5.41 of \cite{van2000asymptotic}. For this proof, we take the working propensity score and the proportion of always-recruited among the always- and incentivized-recruited groups as known. The extension to estimated propensity score and estimated proportion of always-recruited is straightforward, and discussed immediately after the proof. Since we assume monotonicity, the stratum of disincentivized-recruited individuals is empty, and quantities that correspond to $\stratum[] \in \{a, d\}$ are the same as those for $\stratum[] = a$.
	
	\paragraph{The estimating equations}
	
	Let $\weight[\trtr][a](\covsr)$ denote the weight for a unit with $\trt[] = \trtr$ and $\covs[] = \covsr$ for the estimand $\tau^O_a$, $\weight[\trtr][a,c](\covsr)$  the weight for a unit with $\trt[] = \trtr$ and $\covs[] = \covsr$ for the estimand $\tau^O_{a,c}$, and $\weight[\trtr][R](\covsr)$ the weight for a unit with $\trt[] = \trtr$ and $\covs[] = \covsr$ for the estimand $\tau^R$. 
	
	Let $\bm \theta^a = (\theta_1^a, \theta_2^a, \theta_3^a, \theta_4^a)^\top$ and similarly defined vectors of length four for $\bm \theta^{a,c}$ and $\bm \theta^R$. Then, let
	\begin{equation}
	\psi^a(\obsdata; \bm \theta^a) =
	\begin{pmatrix}
	\sum_{i \in j} \enrol \weight[1][a](\covs)\trt\out - \theta_1 \sum_{i \in j} \enrol  \\[5pt]
	\sum_{i \in j} \enrol \weight[0][a](\covs)(1 - \trt)\out - \theta_2 \sum_{i \in j} \enrol \\[5pt]
	\sum_{i \in j} \enrol \weight[1][a](\covs)\trt - \theta_3 \sum_{i \in j} \enrol \\[5pt]
	\sum_{i \in j} \enrol \weight[0][a](\covs)(1 - \trt) - \theta_4 \sum_{i \in j} \enrol
	\end{pmatrix},
	\label{app_eq:estimating_equation}
	\end{equation}
	and similarly defined $\psi^{a,c}(\obsdata; \bm \theta^{a,c})$ and $\psi^{R}(\obsdata; \bm \theta^{R})$. Then, for $\bm \theta = (\bm \theta^a{}^\top, \theta^{a,c}{}^\top, \theta^R{}^\top)^\top$, our estimating equations are of length 12 and defined as
	\[
	\psi(\obsdata; \bm \theta) = \begin{pmatrix}
	\psi^a(\obsdata; \bm \theta^a) \\ \psi^{a,c}(\obsdata; \bm \theta^{a,c}) \\ \psi^R(\obsdata; \bm \theta^R)
	\end{pmatrix}.
	\]
	
	\paragraph{The true solution to the estimating equations}
	
	We use $\psi_k(\obsdata; \bm \theta)$ to denote the $k^{th}$ entry in the vector $\psi(\obsdata; \bm \theta)$. Then, for $\psi_1(\obsdata; \bm \theta)$, it holds that
	\begin{equation}
	\begin{aligned}
	& \E \left[ \sum_{i \in j} \enrol \weight[1][a](\covs)\trt\out - \theta_1^a  \sum_{i \in j} \enrol \right]  = 0 \\
	\iff & 
	\E \left\{ \sum_{i \in j} \E \left[ \enrol \weight[1][a](\covs)\trt\out \mid n_j \right] \right\} = \E[n_j] \theta_1^a
	\end{aligned}
	\label{app_eq:solving_est_eq2}
	\end{equation}
	and
	\begin{align*}
	& \E \left[ \enrol \weight[1][a](\covs)\trt\out  \mid n_j \right] \\
	&= \E \left[ \enrol \weight[1][a](\covs)\trt\out(1)  \mid n_j \right] \\
	&= \E \left[ \enrol \weight[1][a](\covs) \out(1)  \mid \trt = 1, n_j \right] P(\trt = 1 \mid n_j) \\
	&= \E \left\{ \E \left[ \enrol \weight[1][a](\covs) \out(1)  \mid \trt = 1, \out(0), \out(1), \covs \right] \mid \trt = 1, n_j \right\} P(\trt = 1)
	\tag{$\covs$ includes cluster-level information such as $n_j$} \\
	&=  \E \left\{ \weight[1][a](\covs) \out(1) P \left( \enrol = 1 \mid \trt = 1, \out(0), \out(1), \covs \right) \mid \trt = 1, n_j \right\} P(\trt = 1) \\
	&=  \E \left\{ \weight[1][a](\covs) \out(1) P \left( \enrol = 1 \mid \trt = 1, \covs \right) \mid \trt = 1, n_j \right\} P(\trt = 1) \tag{From \cref{ass:enrol_indep}}
	\end{align*}
	
	Then, from \cref{app_eq:1_wps},
	we have that
	\begin{align*}
	\weight[1][a](\covs) & P \left( \enrol = 1 \mid \trt = 1, \covs \right) = \\
	&= \frac{1 - \wps[\covs]}{\wps[\covs]} P \left( \enrol = 1 \mid \trt = 1, \covs \right) \\
	&= \frac{ P(\trt = 0 \mid \covs, \enrol = 1)}
	{\frac{P(\enrol = 1 \mid \trt = 1, \covs)}{\prob(\enrol = 1 \mid \covs)} \prob(\trt = 1)}
	P \left( \enrol = 1 \mid \trt = 1, \covs \right) \\
	&= \frac{ P(\trt = 0 \mid \covs, \enrol = 1) \prob(\enrol = 1 \mid \covs) }
	{\prob(\trt = 1)} \\
	&=  \frac{ P(\enrol = 1 \mid \trt = 0, \covs) \prob(\trt = 0) } {\prob(\trt = 1)}  \\
	&= \frac{P(\enrol(0) = 1 \mid \covs) \prob(\trt = 0) } {\prob(\trt = 1)} \tag{From \cref{app_eq:enrol_given_trt}}.
	\end{align*}
	Returning to the previous quantity we have that
	\begin{align*}
	& \E \left\{ \enrol \weight[1][a](\covs)\trt\out  \mid n_j \right\} = \\
	&= \E \left\{ \out(1)  P(\enrol(0) = 1 \mid \covs)
	\mid n_j \right\}
	\frac{\prob(\trt = 0) } {\prob(\trt = 1)} 
	P(\trt = 1) \\
	&= P(\trt = 0) \E \left\{ \out(1)  h_a^O(\covs) \mid n_j \right\}  \tag{for $g_a^O(\covsr) = f_{\covs[]}(\covsr \mid S = a)$ based on \cref{app_eq:dist_R0_1}}.
	\end{align*}
	Returning back to \cref{app_eq:solving_est_eq2}, we have that
	\begin{align*}
	\E \left\{ \sum_{i \in j} \E \left[ \enrol \weight[1][a](\covs)\trt\out \mid n_j \right] \right\}
	&= P(\trt = 0) \ \E \left\{ n_j \E \left[ \frac1{n_j} \sum_{i \in j}  \out(1)  h_a^O(\covs) \mid n_j \right] \right\} \\
	&= P(\trt = 0) \E(n_j) \E \left[ \frac1{n_j} \sum_{i \in j}  \out(1)  h_a^O(\covs) \right]
	\tag{From \cref{app_ass:regularity}\ref{app_ass_item:n_po_indep}} \\
	&= P(\trt = 0) \E(n_j) \E \left[ \out(1)  h_a^O(\covs) \right].
	\end{align*}
	Following similar steps for the remaining $\theta$s while noting that $E[h_a^O(\covs)] = 1$, we have that the vector $\bm \theta^a_0 = (\theta_{01}^a, \theta_{02}^a, \theta_{03}^a, \theta_{04}^a)^\top$ for
	\begin{align*}
	\theta_{01}^a &= P(\trt = 0) \E \left[ \out(1)  h_a^O(\covs) \right] \\
	\theta_{02}^a &= P(\trt = 0) \E \left[ \out(0)  h_a^O(\covs) \right] \\
	\theta_{03}^a &= P(\trt = 0) \\
	\theta_{04}^a &= P(\trt = 0)
	\end{align*}
	satisfies  that $\E \left\{ \psi^a(\obsdata; \bm \theta_0^a) \right\} = 0$.
	Similarly defined $\bm \theta_0^{a,c}$ and $\bm \theta_0^R$ satisfy that
	$\E \left\{ \psi^{a,c}(\obsdata; \bm \theta_0^{a,c}) \right\} = 0$
	and
	$\E \left\{ \psi^R(\obsdata; \bm \theta_0^R) \right\} = 0$, and as a result
	$\bm \theta_0 = (\bm \theta_0^a{}^\top, \theta_0^{a,c}{}^\top, \theta_0^R{}^\top)^\top$ satisfy that
	$\E \left\{ \psi(\obsdata; \bm \theta_0) \right\} = 0$.
	Furthermore, the causal effects of interest can be written as
	\[
	\effect^O_a = \frac{\theta_{01}^a}{\theta_{03}^a} - \frac{\theta_{02}^a}{\theta_{04}^a}, \quad
	\effect^O_{a,c} = \frac{\theta_{01}^{a,c}}{\theta_{03}^{a,c}} - \frac{\theta_{02}^{a,c}}{\theta_{04}^{a,c}}, \quad \text{and} \quad
	\effect^R = \frac{\theta_{01}^R}{\theta_{03}^R} - \frac{\theta_{02}^R}{\theta_{04}^R}.
	\]

	\paragraph{The empirical solution to the estimating equations}
	
	Next we shift our attention to the empirical version of the estimating equations, and let $\widehat{\bm \theta}^a = (\widehat \theta_1^a, \widehat \theta_2^a, \widehat \theta_3^a, \widehat \theta_4^a)^\top$ denote the solution to $\Psi_J^a(\bm \theta^a) = \sum_{j = 1}^J \psi^a(\obsdata; \bm \theta^a) = 0$. We have that
	\begin{align*}
	\widehat \theta_1^a &= \frac1{\sum_j \sum_{i \in j} \enrol} \sum_j \sum_{i \in j} \enrol \weight[1][a](\covs) \trt\out
	= \frac1n \sum_{i = 1}^n \weight[1][a](\covs) \trt\out \tag{$i = 1, 2, \dots, n$ represent the enrolled individuals in the $J$ clusters}, \\
	\widehat \theta_2^a &= \frac1N \sum_{i = 1}^N \weight[0][a](\covs) (1 - \trt) \out \\
	\widehat \theta_3^a &= \frac1N \sum_{i = 1}^N \weight[1][a](\covs) \trt \\
	\widehat \theta_4^a &= \frac1N \sum_{i = 1}^N \weight[0][a](\covs) (1 - \trt).
	\end{align*}
	We similarly define $\widehat{\bm \theta}^{a,c}$ and $\widehat{\bm \theta}^R$, and $\widehat{\bm \theta}$ is the concatenation of the three vectors. 
	Then, the causal estimators can be written as
	\[
	\widehat \effect_a^O = \frac{\widehat \theta_1^a}{\widehat \theta_3^a} - \frac{\widehat \theta_2^a}{\widehat \theta_4^a}, \quad
	\widehat \effect_{a,c}^O = \frac{\widehat \theta_1^{a,c}}{\widehat \theta_3^{a,c}} - \frac{\widehat \theta_2^{a,c}}{\widehat \theta_4^{a,c}},
	\quad \text{and} \quad
	\widehat \effect^R = \frac{\widehat \theta_1^R}{\widehat \theta_3^R} - \frac{\widehat \theta_2^R}{\widehat \theta_4^R}.
	\]

	\paragraph{Showing that the theorem's conditions hold}
	
	\begin{enumerate}[leftmargin=*,itemsep=5pt]
		\item We show that $\widehat{\bm \theta}$ is consistent for $\bm \theta_0$. For $k, l \in \{ 1, 2, \dots, 12\}$ we have that
		\[
		\frac{\partial}{\partial \theta_l} \psi_k(\obsdata; \bm \kappa) =
		\begin{cases} - \sum_{i \in j} R_i < 0, & \text{if } k = l \\
		0, & \text{otherwise}
		\end{cases}
		\]
		which implies that $\bm \theta_0$ and $\widehat{\bm \theta}$ are both unique roots of the corresponding population and sample level equations (if $\wpspar = \wpspar_0$). The uniqueness of the roots implies that $\widehat{\bm \theta}$ is consistent for $\bm \theta_0$ \citep[Lemma A, Section 7.2.1,][]{serfling2009approximation}.

		\item Since $\psi(\obsdata; \bm \theta)$ is linear in $\theta_k$, for all $k$, we have that $\psi(\obsdata; \bm \theta)$ is twice continuously differentiable as a function of $\bm \theta$.
		
		\item We want to show that $\E\left\| \psi(\obsdata; \bm \theta_0) \right\|^2_2 < \infty$. Take the first entry of the vector:
		\begin{align*}
		\E \left[ \psi_1(\obsdata; \bm \theta_0)^2 \right]
		&= \E \left[ \Big| \sum_{i \in j} \enrol \weight[1][a](\covs)\trt\out - \theta_1 \Big|^2  \right] \\
		&\leq \E \left\{ \left[ \sum_{i \in j} |\enrol \weight[1][a](\covs)\trt\out| + |\theta_1| \right] ^2 \right\}  \\
		&\leq \E \left\{ \left[ \sum_{i \in j} \weight[1][a](\covs) |\out| + |\theta_1| \right] ^2  \right\} \\
		& < c,
		\end{align*}
		for some constant $c$ because $\weight[1][a](\covsr)$ are bounded since $\wps$ is bounded from \cref{app_prop:positivity}, and the outcome and cluster size $N_j$ are bounded with probability 1 from \cref{app_ass:regularity}.
		Similarly we can show that $\E \left[ \psi_k(\obsdata; \bm \theta_0)^2 \right]$ are bounded for all $k$. 
		Combined, these imply that $\E\left\| \psi(\obsdata; \bm \theta_0) \right\|^2_2 < \infty$. 
		
		\item 
		From
		\[
		\frac{\partial}{\partial \theta_l} \psi_k(\obsdata; \bm \kappa) =
		\begin{cases} - \sum_{i \in j} R_i < 0, & \text{if } k = l \\
		0, & \text{otherwise}
		\end{cases}
		\]
		for $k, l$, we have that
		\[
		\E \left[ \frac{\partial}{\partial \bm \theta^T} \psi(\obsdata; \bm \theta) \Big|_{\bm \theta_0} \right] = - E[n_j] I_{12},
		\]
		exists and is non-singular since $N_j$ and therefore $n_j$ is bounded from \cref{app_ass:regularity}.

		\item Since all second-order derivatives of the estimating function are equal to 0, the integrable function $\overset{\cdot\cdot}{\psi}(\obsdata[j][F] = 0$ dominates all second-order partial derivatives
		for all $\bm \theta$ in a neighborhood of $\bm \theta_0$.
	\end{enumerate}

	From Theorem 5.41 of \cite{van2000asymptotic}, we have that
	\[
	\sqrt{J} \left( \widehat{\bm \theta} - \bm \theta_0 \right) \rightarrow N(0, \Sigma(\bm \theta_0) ) \quad \text{as} \quad J \rightarrow \infty,
	\]
	for $\Sigma(\bm \theta_0) = A(\bm \theta_0)^{-1} V(\bm \theta_0) [A(\bm \theta_0)^{-1}]^T$, where
	\[
	A(\bm \theta_0) = \E \left[ - \frac{\partial}{\partial \bm \theta^T} \psi(\obsdata; \bm \theta) \Big|_{\bm \theta_0} \right] = E(-n_j) I_{12}, \quad \text{and} \quad
	V(\bm \theta_0) = \E \left[  \psi(\obsdata; \bm \theta_0)  \psi(\obsdata; \bm \theta_0)^T \right].
	\]
	Consider the multivariate function $g(\bm \theta) = g(\bm \theta^a, \bm \theta^{a,c}, \bm \theta^R)$ defined as
	\[
	g(\bm \theta) =
	\begin{pmatrix}
	\dfrac{\theta_1^a}{\theta_3^a} - \dfrac{\theta_2^a}{\theta_4^a} \\[10pt]
	\zeta_1 \left( \dfrac{\theta_1^{a,c}}{\theta_3^{a,c}} - \dfrac{\theta_2^{a,c}}{\theta_4^{a,c}} \right) + \zeta_2 \left( \dfrac{\theta_1^a}{\theta_3^a} - \dfrac{\theta_2^a}{\theta_4^a} \right) \\[10pt]
	\dfrac{\theta_1^R}{\theta_3^R} - \dfrac{\theta_2^R}{\theta_4^R}
	\end{pmatrix},
	\]
	for constants $\zeta_1 = [ 1 - \pi_a / (\pi_a + \pi_c) ]^{-1}$ and $\zeta_2 = - \zeta_1 (1 - \zeta_1^{-1})$. Then, $g(\widehat {\bm \theta}) = (\widehat \tau^O_a, \widehat \tau^O_c, \widehat \tau^R)^\top$, where $\widehat \tau^O_c$ is calculated using the true proportion of always-recruited among the always- and incentivized-recruited, $\pi_a / (\pi_a + \pi_c)$, and
	$g(\bm \theta_0) = (\tau^O_a, \tau^O_c, \tau^R)^\top$.
	Using the delta method,
	\[
	\sqrt{J} \left( (\widehat \tau^O_a, \widehat \tau^O_c, \widehat \tau^R)^\top - (\tau^O_a, \tau^O_c, \tau^R)^\top \right) \rightarrow N(0, S),
	\]
	where
	\( \displaystyle
	S = (\nabla g(\bm \theta_0))^T \Sigma(\bm \theta_0) \nabla g(\bm \theta_0)
	\)
	and \( \nabla g(\bm \theta_0) = \)
	\[
	\begin{pmatrix}
	\dfrac1{\theta_{03}^a} & - \dfrac{1}{\theta_{04}^a} & - \dfrac{\theta_{01}^a}{\theta_{03}^a{}^2} & \dfrac{\theta_{02}^a}{\theta_{04}^a{}^2} & 0 & 0 & 0 & 0 & 0 & 0 & 0 & 0 \\[10pt]
	\dfrac{\zeta_2}{\theta_{03}^a} & - \dfrac{\zeta_2}{\theta_{04}^a} & -  \dfrac{\zeta_2 \theta_{01}^a}{\theta_{03}^a{}^2} & \dfrac{\zeta_2 \theta_{02}^a}{\theta_{04}^a{}^2} & 
	\dfrac{\zeta_1}{\theta_{03}^{a,c}} & - \dfrac{\zeta_1}{\theta_{04}^{a,c}} & - \dfrac{\zeta_1 \theta_{01}^{a,c}}{\theta_{03}^{a,c}{}^2} & \dfrac{\zeta_1 \theta_{02}^{a,c}}{\theta_{04}^{a,c}{}^2}
	& 0 & 0 & 0 & 0 \\[10pt]
	0 & 0 & 0 & 0 & 0 & 0 & 0 & 0 & \dfrac1{\theta_{03}^R} & - \dfrac{1}{\theta_{04}^R} & - \dfrac{\theta_{01}^R}{\theta_{03}^R{}^2} & \dfrac{\theta_{02}^R}{\theta_{04}^R{}^2}
	\end{pmatrix}.
	\]
\end{proof}

\paragraph{Extensions to the asymptotic results}

First, the asymptotic results of consistency and asymptotic normality can be extended to accommodate the estimated propensity score from a correctly specified parametric model. If $\alpha$ is the vector of parameters, one needs to extend the existing estimating equations in our proof to include the derivative of the log pseudo-likelihood contribution for cluster $j$, $\psi_\alpha(\obsdata; \alpha) = \partial \left\{ \sum_{i \in j} R_{ij} \log \left[ \wpsmod ^{\trt} (1 - \wpsmod)^{1 - \trt} \right] \right\} / \partial \alpha$.

Secondly, the results extend to the case where $\pi_a / (\pi_a + \pi_c)$ is estimated. First, in randomized experiments, the randomization probability $\pi^t$ is most often known, and the proportion of treated individuals in the recruited population is estimated by its empirical version. Therefore, we estimate
\[
\widehat{\frac{\pi_a}{\pi_a + \pi_c}} = \frac{\pi^t}{1 - \pi^t} \frac{1 - \widehat p^t}{\widehat p^t} 
= \frac{\pi^t}{1 - \pi^t} \frac{\sum_j \sum_{i \in j}\enrol(1 - \trt)}{\sum_j \sum_{i \in j} \enrol \trt}.
\]
Therefore, we can acquire consistency and asymptotic normality of our estimators when the proportion of always-recruited in the always- and incentivized-recruited groups is estimated by extending the estimating equation to include terms $\sum_{i\in j} \enrol (1 - \trt)$ and $ \sum_{i \in j} \enrol \trt$.

\subsection{Sensitivity analysis}
\label{supp_subsec:sensitivity}

\begin{proof}[Proof of \cref{prop:sensitivity}]
	We consider $\Gamma$-sized deviations to the enrollment process due to an unmeasured variable expressed as
	\(
	\Gamma^{-1} \leq {\delta(\covsr)} \ / \ {\delta^*(\covsr, u)} \leq \Gamma.
	\)
	Note also, that, using that $\wps = {\delta(\covsr)}/\{\delta(\covsr) + r^{-1}\} $, and the definitions of $\weight[1ij], \weight[1ij][*]$
	\[
	\rho_{ij} = 
	\frac{\weight[1ij][*]}{\weight[1ij]} =
	\frac{1 - \wps[\covsr,u][*]}{\wps[\covsr,u][*]}
	\frac{\wps}{1 - \wps} 
	= \dfrac{r \delta(\covsr)}{r \delta^*(\covsr, u)} = \dfrac{\delta(\covsr)}{\delta^*(\covsr, u)} \in [\Gamma^{-1}, \Gamma].
	\]
	Using $\rho_{ij}$ and $\weight[1ij]$, the estimator which uses the true propensity score and we wish to bound can be written as
	\begin{align*}
	\frac{\sum_{ij} \weight[1ij][*] \trt \out}
	{\sum_{ij} \weight[1ij][*] \trt} 
	&= \frac{\sum_{ij}
		\rho_{ij} \weight[1ij] \trt \out}
	{\sum_{ij} \rho_{ij} \weight[1ij] \trt}.
	\end{align*}
	The problem of maximizing (or minimizing) the last quantity under the constraints $\rho_{ij} \in [\Gamma^{-1}, \Gamma^1]$ can be transformed to a linear programming problem using the Charnes-Cooper transformation \citep{charnes1962programming}. The linear program under this transformation is the one in the main text, where $\lambda_{ij} = \kappa \rho_{ij}$.
\end{proof}

Bounds for the causal effect estimator on the always- and incentivized-recruited individuals, $\widehat \effect^O_{a,c}$, are acquired similarly.
For this estimator, the weights of the treated individuals are equal to 1, and we can focus on bounding the part of the estimator that corresponds to the control individuals.
Denote the weights for the controls as $w_{0ij}^* = w_0^*(X_{ij}, U_{ij}) = e^*(X_{ij}, U_{ij}) / [1 - e^*(X_{ij}, U_{ij})]$. The following proposition provides an algorithm to acquire the bounds of the estimator for violations of the ignorable recruitment assumption up to $\Gamma$.


\begin{proposition}
	Maximizing (minimizing) $ \tau^*_0 = \frac{\sum_{ij} w_{0ij}^* (1 - \trt) \out}{\sum_{ij} w_{0ij}^* (1 - \trt)} $ under the $\Gamma-$violation of the ignorable recruitment assumption is equivalent to solving the linear program that maximizes (minimizing)
	\(
	\sum_{ij} \lambda_{ij} \weight[0ij] (1 - \trt) \out 
	\) 
	with respect to $\lambda_{ij}$ subject to three constraints:
	\begin{enumerate*}[label = (\alph*)]
		\item 
		\( {\kappa}{\Gamma^{-1}} \leq \lambda_{ij} \leq \kappa \Gamma \)
		\item 
		\(
		\sum_{ij} \lambda_{ij} \weight[0ij] (1 - \trt) = 1,
		\)
		and
		\item
		\(
		\kappa \geq 0,
		\)
	\end{enumerate*}
	where $\weight[0ij] = w_0(X_{ij})$ is the weight of unit $i$ under control and working propensity score $e(x)$, and $\kappa$ is a parameter of the linear program.
\end{proposition}

\begin{proof}
	The proof is very similar to that of \cref{prop:sensitivity}.
\end{proof}

These propositions allow us to bound the Haj\'ek estimators $\widehat \effect^O_a$ and $\widehat \effect^O_{a,c}$ under $\Gamma$-violations of the ignorable recruitment assumption by solving a linear program, which can be achieved computationally fast. On the other hand, the causal estimator for the effect on the incentivized-recruited group is equal to
\[
\widehat \effect^O_c = 
\frac{\pi_a + \pi_c}{\pi_c} \ \widehat \effect^O_{a,c} -
\frac{\pi_a}{\pi_c} \ \widehat \effect^O_a.
\]
The optimization problem for bounding this quantity under $\Gamma$-violation of the ignorable recruitment assumption cannot be immediately written as a linear program, and it is likely to require computationally intensive optimization tools. Instead, we use the bounds acquired for $\widehat \effect^O_a$ and $\widehat \effect^O_{a,c}$ to also acquire bounds for $\widehat \effect^O_c$. Specifically, let $l_a, u_a$, and $l_{a,c}, u_{a,c}$ be the lower and upper bounds for $\widehat \effect^O_a$ and $\widehat \effect^O_{a,c}$, respectively. Then,
we use $l_c = \frac{\pi_a + \pi_c}{\pi_c} \ l_{a,c} - \frac{\pi_a}{\pi_c} u_a$ and $u_c = \frac{\pi_a + \pi_c}{\pi_c} \ u_{a,c} - \frac{\pi_a}{\pi_c} l_a$ as the lower and upper bounds for $\widehat \effect^O_c$, respectively. If $l_c^*$ and $u_c^*$ denote the theoretical lower and upper bounds for $\widehat \effect^O_c$, then $[l_c^*, u_c^*] \subseteq [l_c, u_c]$. Therefore, our bounds are valid in that the causal estimator falls within the designed interval under a $\Gamma$-violation of the ignorable recruitment assumption. However, this interval might be wider than necessary. The proof is straightforward, hence omitted.

\section{Simulations}
\label{supp_sec:sims}

\subsection{Data generative mechanisms}
\label{supp_subsec:sim_dgm}

We consider 36 data generative mechanisms that are combinations of the specifications described in \cref{tab:sim_setup}. Here, we provide the details for how the data are generated. We considered three choices for the number of clusters where $J \in \{200, 500, 800\}$. Across all scenarios we specified that each cluster had 100 individuals in its overall population, $N_j = 100$, though (as described below) at least 50\% were never-recruited individuals, and therefore $n_j$ was in general smaller than 50.

\begin{enumerate}[leftmargin=*,itemsep=0pt]
	\item Cluster treatment was generated by choosing $m$ out of $J$ clusters to receive the treatment, where we used $m / J \in \{0.25, 0.5\}$ representing an imbalanced and a balanced treatment assignment, respectively.
	
	\item We generated three individual level covariates $\covss$ and two cluster level covariates $\covsc$, with $\covs = (\covss, \covsc)$. Covariates $\covss[1ij], \covss[3ij], \covsc[1j]$ were generated from independent standard normal distributions truncated to lie between -3 and 3, and $\covss[2ij], \covsc[2j]$ were generated from independent Bernoulli distributions with probability of success 0.5.
	
	\item We considered outcomes with intra-class correlation. Specifically, we generated cluster-level random effects $\epsilon_j^{icc}$ from a $N(0, \sigma^2 \rho)$ distribution. Residual variability was generated as $
	\epsilon_{ij}$ from a $N(0, \sigma^2 (1- \rho))$ distribution Then, potential outcomes were calculated as
	\[
	Y_{ij}(\trtr) = \covs^T \beta_{\trtr}^{\out[]} + \epsilon_{ij} + \epsilon_j^{icc}.
	\]
	
	\item The potential enrollment under control $\enrol(0)$ was generated from a Bernoulli distribution with a logistic link function and linear predictor $\covs^T \beta^R_0$. 
	
	\item Under monotonicity, the potential enrollment values under treatment are generated conditional on the potential enrollment values under control. We set  $\enrol(1) = 1$ for those with $\enrol(0) = 1$. We consider parameters $\wpspar$ and $\delta(x; \wpspar) = 1 + \exp \{ - x^t \wpspar \}$.
	Then, for a unit with $\enrol(0) = 1$ we generate the corresponding $\enrol(1)$ from a Bernoulli distribution with probability of success $[\delta(\covs; \wpspar) - 1] \exp \{ \covs^T \beta^R_0 \} $. Doing so ensures that $\delta(\covsr; \wpspar)$ is in fact equal to the ratio in \cref{ass:po_diffenrol} which is a ratio of marginal probabilities for the recruitment under the two treatments.
	
\end{enumerate}

We considered data generative mechanisms that led to different prevalence of the principal strata. We named these Scenarios A, B, and C. The proportions of the principal strata are shown on \cref{tab:sim_setup}. We also considered sub-cases that varied how strongly the covariates separate the always-recruited from the group of always- and incentivized-recruited. We named these Cases 1 and 2. We did this by varying the size of the coefficients for the covariates in the model for $\delta(x)$ since
\[
\delta(x)^{-1} = P(S = a \mid S \in \{a, c \}, X).
\]
Therefore, larger (in absolute value) coefficients for $x$ in $\delta(x)$ lead to stronger covariate separation between the always and the incentivized-recruited.

We found parameters that achieve these specifications by noting that increasing the intercept in $\delta(\covsr; \wpspar)$ does not affect the proportion of the population that is always-recruited, whereas it decreases the proportion of incentivized-recruited and increases the proportion of never-recruited. Also, we noted that increasing the intercept in the model for $\enrol[](0)$ (the intercept in $\beta^R_0$) reduces the number of never recruited, but does not affect the proportion of always-recruited among the always- and incentivized-recruited. 

We found that the coefficients listed in \cref{tab:sim_coefs} achieve these goals in terms of principal strata distribution.
We also set:
\begin{enumerate*}[label=(\alph*)]
	\item $ \beta_{1}^{\out[]}  = (2, - 1, 3, 0.1, - 0.1, 0.3)$,
	\item $ \beta_{0}^{\out[]} = (0, - 0.5, 1, 0.1, - 0.2, 0.3)$,
	\item $\sigma^2 = 1$, and
	\item $\rho = 0.1$.
\end{enumerate*}

\begin{table}[!t]
	\centering
	\caption{Coefficients in the data generative mechanisms in the simulations. These coefficients are chosen to achieve target principal strata distribution in the overall population and vary the covariate separation among always- and incentivized-recruited.}
	\label{tab:sim_coefs}
	\vspace{5pt}
	\begin{tabular}{cccc}
		\hline \\[-10pt]
		& $\beta_0^R$ &  & $\wpspar$ \\
		\cmidrule(lr){2-2} \cmidrule(lr){4-4} 
		\multirow{2}{*}{Scenario A} & \multirow{2}{*}{(-0.99, 0.3, -0.6, 0, 0.1, -0.3)} &Case 1& (0.275, 0.3, -0.5, -0.1, 0, -0.15) \\
		& &Case 2 & (0.160, 0.2, -0.3, -0.1, 0, -0.15) \\
		\multirow{2}{*}{Scenario B} & \multirow{2}{*}{(-0.70, 0.3, -0.6, 0, 0.1, -0.3)} & Case 1 & (0.275, 0.3, -0.5, -0.1, 0, -0.15) \\
		& &Case 2 &  (0.160, 0.2, -0.3, -0.1, 0, -0.15) \\
		\multirow{2}{*}{Scenario C} & \multirow{2}{*}{(-1.35, 0.3, -0.6, 0, 0.1, -0.3)} & Case 1 & (-0.235, 0.3, -0.5, -0.1, 0, -0.15) \\
		&  &Case 2 &  (-0.340, 0.2, -0.3, -0.1, 0, -0.15) \\
		\hline
	\end{tabular}
\end{table}

\subsection{ Additional results}
\label{supp_subsec:add_sim_results}

\begin{figure}[!t]
	\centering
	\includegraphics[width=0.9\textwidth]{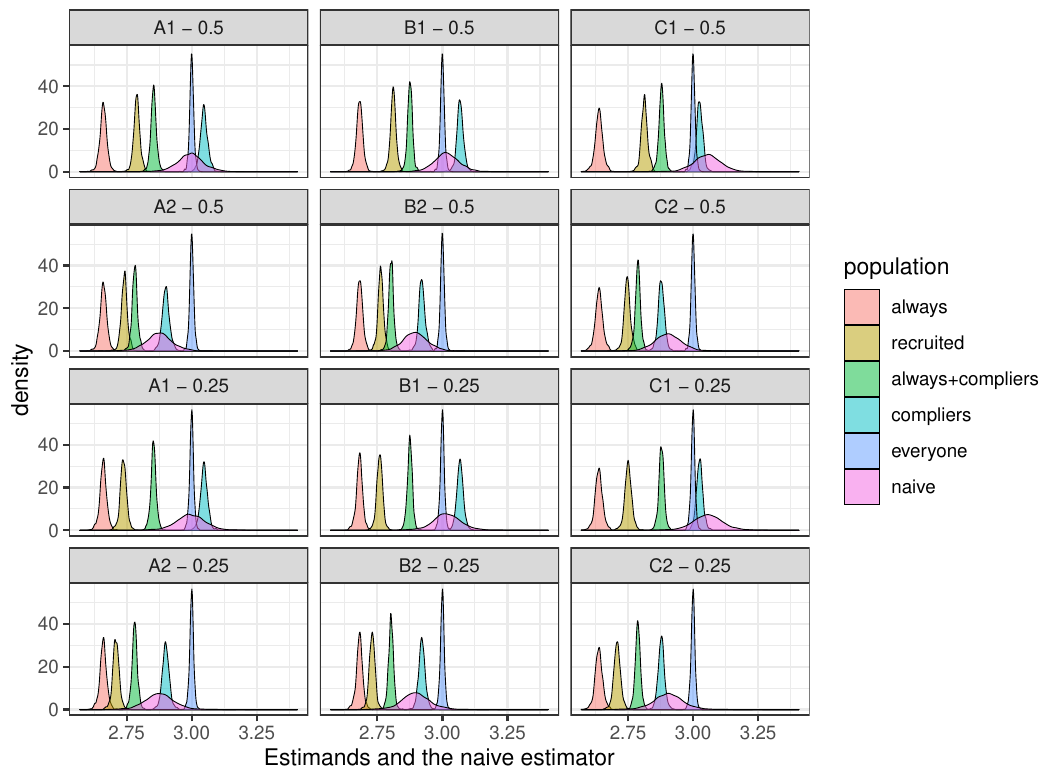}
	\caption{Distribution of sample causal effect among different populations and estimated na\"ive difference of mean outcomes across 500 simulated data sets and number of clusters $J \in \{200, 500, 800\}$ organized by the 12 combinations of scenario, case, and treatment proportion in our configurations.}
	\label{supp_fig:sim_naive_estimator}
\end{figure}

\cref{supp_fig:sim_naive_estimator} shows the distribution of treatment effects among the different populations of interest (everyone in the overall population, the recruited population, always- and disincentivized-recruited strata, always- and incentivized- recruited strata, and the incentivized-recruited strata) and the distribution for the difference of mean outcomes among treated and control recruited individuals. The latter quantity corresponds to the na\"ive estimator that ignores recruitment bias. The na\"ive estimator does {\it not} return quantities that represent a causal effect over an interpretable population.

\begin{figure}[!t]
	\centering
	\includegraphics[width=0.95\textwidth]{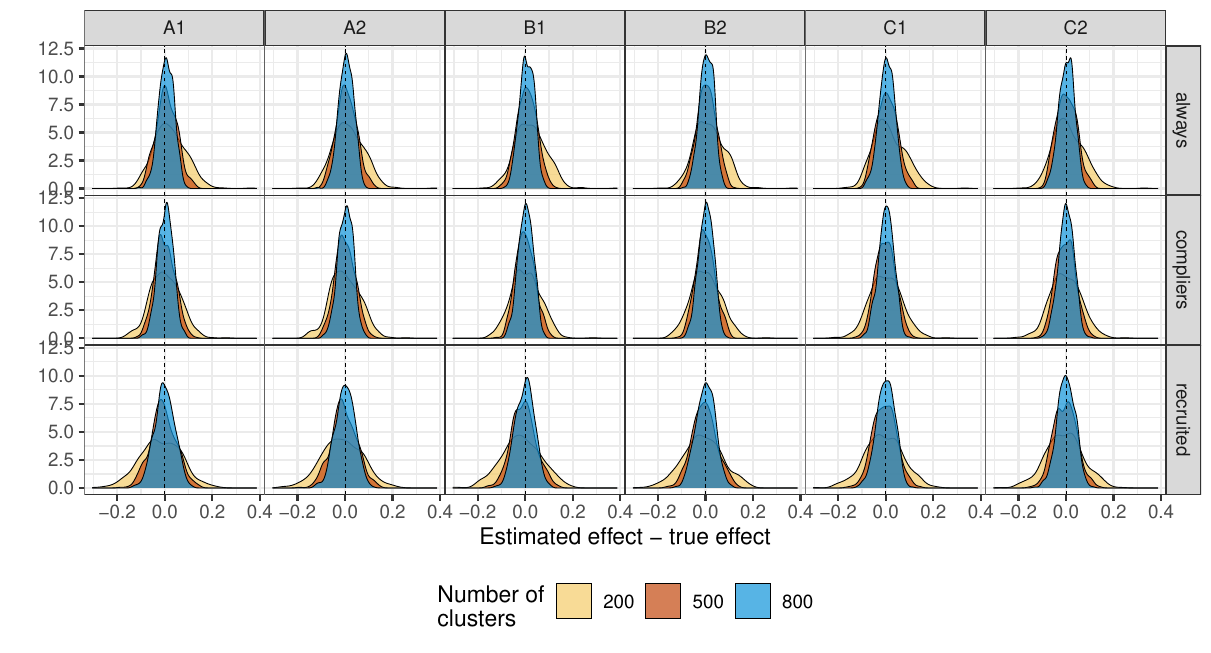}
	\caption{Bias of the causal estimator for the treatment effect among the always-recruited, the incentivized-recruited, and the recruited populations, across 500 data sets and under 6 scenarios and 3 choices for the number of clusters, when the probability of cluster treatment is 0.25.}
	\label{supp_fig:bias_estPS_r13}
\end{figure}

\begin{figure}[!t]
	\includegraphics[width=\textwidth]{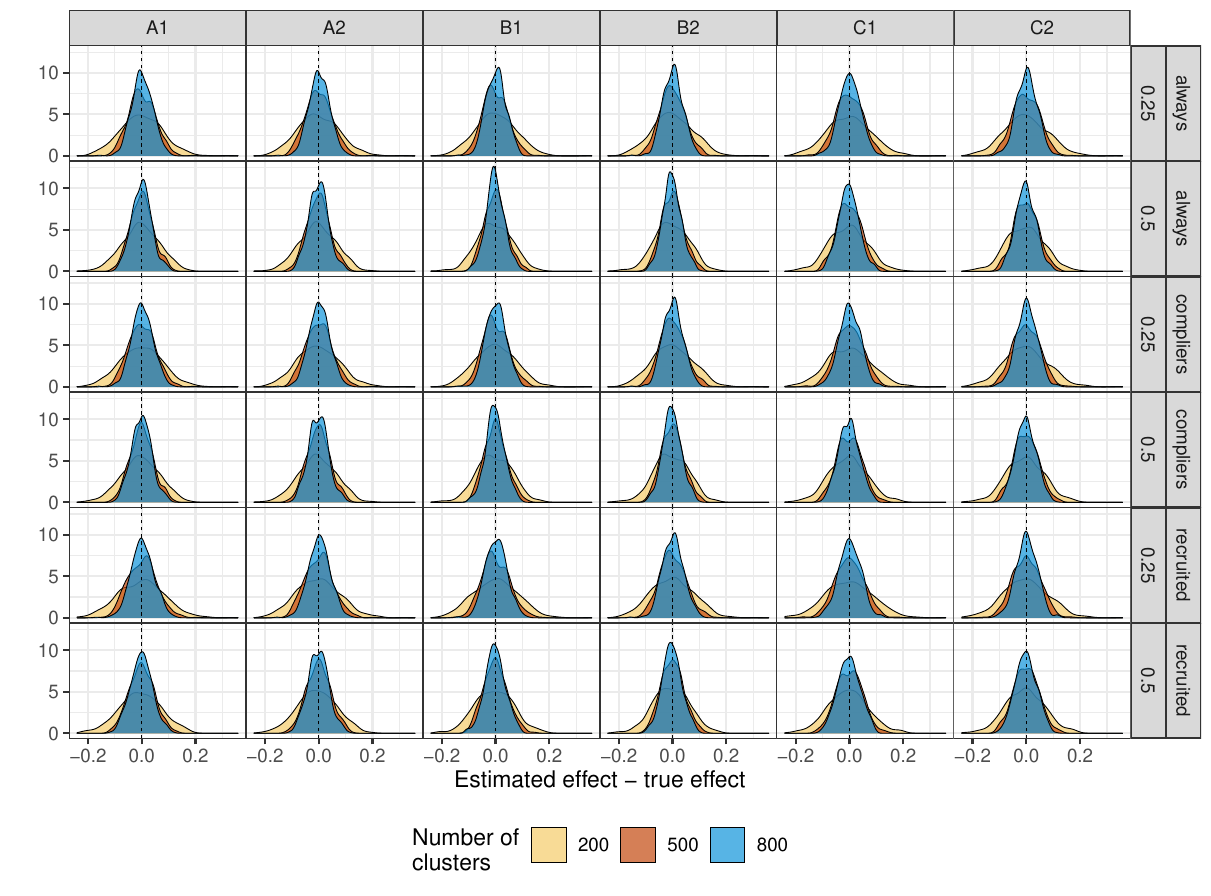}
	\caption{Bias of the causal estimator that uses the known working propensity score for the treatment effect among the always-recruited, the incentivized-recruited, and the recruited populations, across 500 data sets and under 6 configurations, 3 choices for the number of clusters, and treatment of clusters set to 0.25 or 0.5.}
	\label{supp_fig:bias_knownPS}
\end{figure}

\cref{supp_fig:bias_estPS_r13} shows the estimated minus the true causal effect on the the three populations of interest, the recruited population, the always-recruited and the incentivized-recruited population, when using the proposed estimators with estimated working propensity score and imbalanced designs that assign 25\% of the clusters to treatment. These results show that the estimators are essentially unbiased across all scenarios considered. Similarly, \cref{supp_fig:bias_knownPS} shows the same quantity across all simulation configurations and when using the known working propensity score. As expected, the estimator of the causal effect in the three populations is unbiased, and it is more precise under a larger number of clusters.

\begin{figure}[!t]
	\centering
	\includegraphics[width=0.98\textwidth]{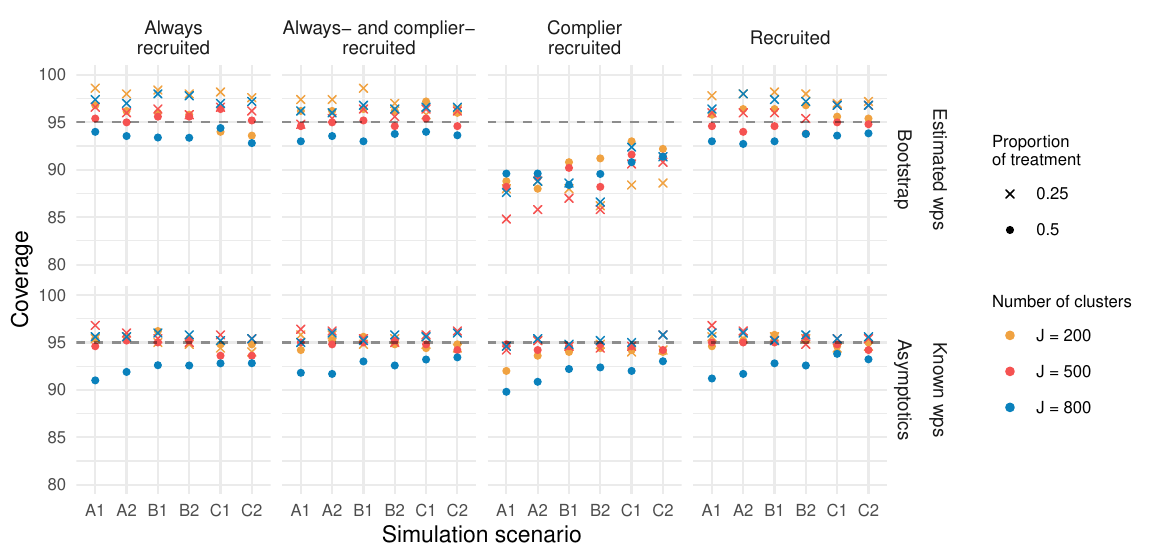}
	\caption{Coverage of 95\% intervals for the estimator of the effect on the recruited, the always-recruited, and the incentivized-recruited populations, based on the asymptotic distribution for the known working propensity score and the bootstrap for estimated propensity score.}
	\label{fig:sim_coverage}
\end{figure}

Also, \cref{fig:sim_coverage} shows the coverage of the 95\% confidence intervals for the causal estimators for the effect on the recruited, the always-recruited, the combination of always- and incentivized-recruited, and the incentivized-recruited populations. For the estimators that use the known propensity score, we consider 95\% intervals based on the derived asymptotic distribution. For the estimators that use the estimated propensity score, we consider a bootstrap procedure that resamples clusters while holding the number of treated clusters fixed, in order to emulate cluster treatment assignment under the simulation design. We calculate the standard deviation of the bootstrap estimates and use them to construct 95\% confidence intervals. The coverage of the intervals based on the asymptotic distribution for the estimator that uses the known propensity score is close to 95\% across all scenarios. Coverage based on the bootstrap is close to 95\% for the estimator of the causal effect on the recruited population, the always-recruited individuals in the overall population, and the combination of always- and incentivized-recruited individuals in the overall population when using the estimated propensity score. Coverage of the same intervals for the causal effect on the incentivized-recruited group are lower, ranging from 85 to 93\%, though always closer to the nominal level under Scenario C compared to Scenarios A and B.

\begin{figure}[!t]
	\centering
	\includegraphics[width=0.98\textwidth]{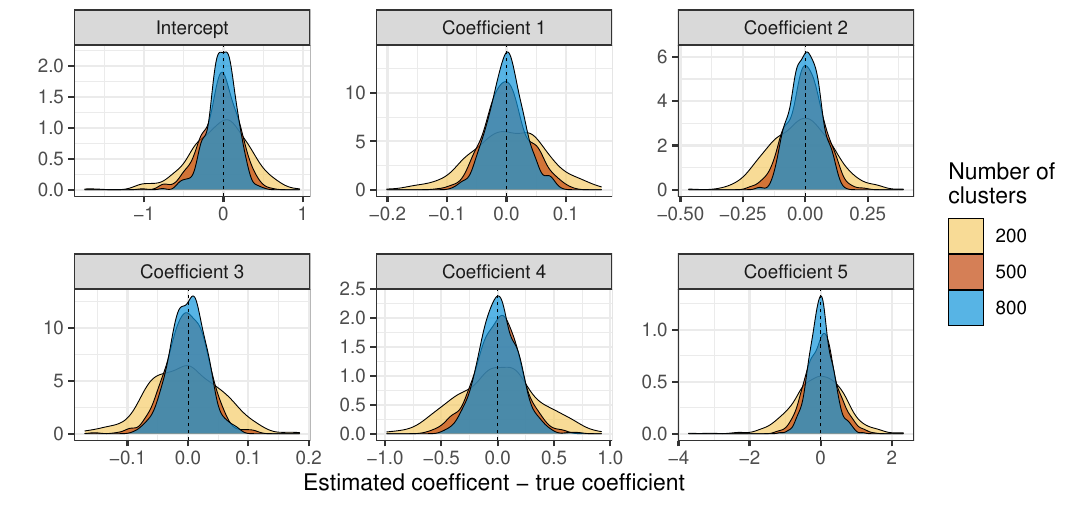}
	\caption{Distribution of estimated minus true working propensity score coefficients for scenario A, case 1 across simulated data sets, and for different number of clusters.}
	\label{fig:sim_wps_A1}
\end{figure}

Lastly, we investigate the estimation technique of our working propensity score model parameters. In \cref{fig:sim_wps_A1} we show the distribution of estimated minus true parameter across the simulated data sets for scenario A, case 1, and under different number of clusters. Results from the remaining simulation configurations were similar. We see that the working propensity score estimation technique based on the pseudo-likelihood returns unbiased estimates of the working propensity score parameters, that are closer concentrated near the true value for a larger number of clusters.

\section{Simulations under violation of \cref{ass:enrol_indep}}
\label{supp_sec:sims_NDR}

In \cref{prop:comparing_assumptions}, we established that \cref{ass:enrol_indep} is equivalent to $S \indep \{Y(0), Y(1)\} \mid X$. Then, in \cref{prop:comparing_assumptions2}, we showed that \cref{ass:po_diffenrol} and \cref{ass:monotonicity} imply that $S \indep \{Y(0), Y(1)\} \mid X, S \in \{a, c\}$.
Since (under monotonicity) the recruited group corresponds to always- and incentivized-recruited individuals only, it is reasonable to wonder whether \cref{ass:po_diffenrol} and \cref{ass:monotonicity} suffice for identifying overall causal effects.

We performed simulations to investigate this. These simulations mimic the scenarios in the simulations of \cref{sec:simulation} detailed in Supplement \ref{supp_sec:sims}, except they are designed to represent violations of the ignorable recruitment \cref{ass:enrol_indep}, while \cref{ass:po_diffenrol} and \cref{ass:monotonicity} still hold.

Specifically, data are generated under all 36 data generative mechanisms detailed in Supplement \ref{supp_subsec:sim_dgm}. The main difference lies on the generation of $\enrol(0)$ which is now allowed to depend on the values of the potential outcome. Specifically, $\enrol$ is generated from a Bernoulli distribution with a logistic link function and linear predictor $(\covs^\top \ \out(0) \ \out(1))^\top \widetilde \beta_0^{\enrol[]}$. \cref{ass:po_diffenrol} still holds, meaning that the ratio of recruitment under different treatment vectors still only depends on measured covariates $X$ and not potential outcomes $Y(0), Y(1)$, therefore it can still be denoted as $\delta(x)$. We ensured the same prevalence of principal strata and treatment in the recruited group for the simulations where \cref{ass:enrol_indep} holds (\cref{sec:simulation}) and for the simulations where \cref{ass:enrol_indep} does not hold. To do so, we had to slightly alter the intercepts for the models used to generating the recruitment indicators (intercept in $\widetilde \beta_0^R$ and $\alpha$) while the coefficients of all measured covariates were kept the same in the two scenarios. The new coefficients are shown in \cref{tab:sim_coefs_NDR}. Everything else remained the same.

\begin{table}[!t]
	\centering
	\caption{Coefficients in the data generative mechanisms in the simulations under violations of the ignorable recruitment assumption. These coefficients are chosen to achieve similar data structure to the ones in the simulations of \cref{sec:simulation}. The last two entries in $\widetilde \beta_0^R$ correspond to the potential outcomes.}
	\label{tab:sim_coefs_NDR}
	\vspace{5pt}
	\begin{tabular}{cccc}
		\hline \\[-10pt]
		& $\widetilde \beta_0^R$ &  & $\wpspar$ \\
		\cmidrule(lr){2-2} \cmidrule(lr){4-4} 
		\multirow{2}{*}{Scenario A} & \multirow{2}{*}{(-0.7 , 0.3, -0.6, 0, 0.1, -0.3, 0.2, -0.12)} & Case 1& (0.270, 0.3, -0.5, -0.1, 0, -0.15) \\
		& &Case 2 & (0.160, 0.2, -0.3, -0.1, 0, -0.15) \\
		\multirow{2}{*}{Scenario B} & \multirow{2}{*}{(-0.4 , 0.3, -0.6, 0, 0.1, -0.3, 0.2, -0.12)} & Case 1 & (0.272, 0.3, -0.5, -0.1, 0, -0.15) \\
		& &Case 2 &  (0.177, 0.2, -0.3, -0.1, 0, -0.15) \\
		\multirow{2}{*}{Scenario C} & \multirow{2}{*}{(- 1.07, 0.3, -0.6, 0, 0.1, -0.3, 0.2, -0.12)} & Case 1 & (-0.268, 0.3, -0.5, -0.1, 0, -0.15) \\
		&  &Case 2 &  (-0.355, 0.2, -0.3, -0.1, 0, -0.15) \\
		\hline
	\end{tabular}
\end{table}

\begin{figure}
	\centering
	\includegraphics[width=\textwidth]{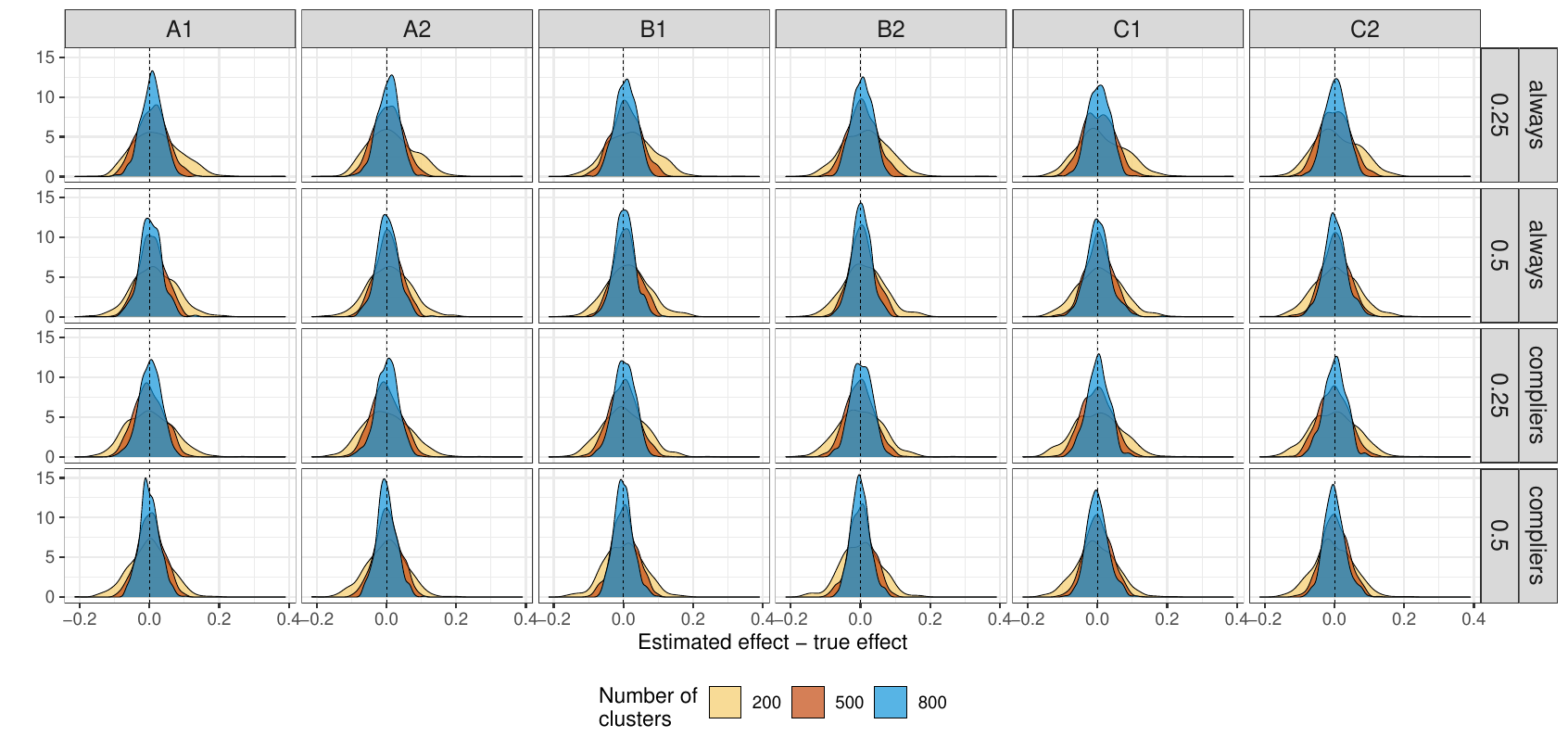}
	\caption{Bias of the causal estimator that uses the estimated working propensity score under violation of the ignorable recruitment assumption. We show results for the treatment effect among the always-recruited and the incentivized-recruited across 500 data sets and under 6 configurations, 3 choices for the number of clusters, and treatment of clusters set to 0.25 or 0.5.}
	\label{supp_fig:sim_results_NDR}
\end{figure}

We show the bias of the proposed estimator based on the estimated working propensity score in \cref{supp_fig:sim_results_NDR}. We exclude estimates for the recruited population since identifiability of such effects only requires \cref{ass:po_diffenrol} that holds. Instead, we focus on the treatment effect on the always-recruited and the incentivized-recruited groups, since these correspond to causal effects over subgroups of the overall population. We find that our causal estimators are essentially unbiased irrespective of the number of clusters, treatment prevalence, and scenario considered. This provides some indication that the assumptions of the non-differential recruitment (\cref{ass:po_diffenrol}) and monotonicity (\cref{ass:monotonicity}) might suffice for identification of the effects in the always- and incentivized-recruited populations. Even though we considered a large number of simulation scenarios, simulations cannot unequivocally establish that identifiability holds, and further research is required.

\section{Information on the ARTEMIS trial}
\label{supp_sec:application}

\cref{tab:ARTEMIS-covariates} provides descriptive information for the enrolled group of patients with treatment (intervention) and without (control). We perform t-tests for continuous covariates and  $\chi^2$-tests for binary covariates to compare the mean of each covariate in the intervention and control groups. We find that multiple covariates have p-values for the corresponding test that are below 0.05, indicating that the treated and control enrolled groups are different with respect to these characteristics.

\begin{table}[!b]
	\centering
	\begin{tabular}{c c c c}
		Covariate  &Intervention & Control & $p$ value \\
		\hline
		Age (year) & 62.0 (11.7) & 61.4 (11.5) & 0.013\\
		Male (\%) & 68.9 (46.3) & 68.3 (46.5) & 0.509\\
		White (\%) & 90.0 (30.1) & 86.4 (34.3) & 0.000\\
		Black (\%) & 8.5 (27.8) & 11.3 (31.6) & 0.000\\
		College (\%) & 49.4 (50.0) & 53.5 (49.9) & 0.000\\
		Private insurance (\%) & 63.2 (48.2) & 65.5 (47.5) & 0.018\\
		Prior MI (\%) & 19.3 (39.5) & 21.3 (41.0) & 0.015\\
		Prior P2Y$_{12}$ use (\%) & 12.6 (33.2) & 16.3 (36.9) & 0.000\\
		Hemoglobin level & 12.87 (2.02) & 12.80 (2.08) & 0.073\\
		Hypertension (\%) & 67.2 (47.0) & 70.9 (45.4) & 0.000\\
		DES use (\%) & 82.8 (37.8) & 78.6 (41.0) & 0.000\\
		Diabetes (\%) & 31.1 (46.3) & 34.3 (47.5) & 0.000\\
		Employment (\%) & 48.3 (50.0) & 46.9 (49.9) & 0.180\\
		Multivessel disease (\%) & 47.1 (49.9) & 45.5 (49.8) & 0.133\\
		CABG (\%) & 1.5 (12.0) & 1.4 (11.6) & 0.800\\
		\hline
	\end{tabular}
	\caption{Mean Baseline Characteristics of Enrolled Patients by Randomized Arm, with standard deviation in brackets. The $p$-value is computed by $t$-test for continuous data and Pearson's chi-squared test for binary data.}
	\label{tab:ARTEMIS-covariates}
\end{table}





\clearpage
\bibliographystyle{plainnat}
\bibliography{CRT_bias,ClusterRandomized,GeneralCausal,Principal-Stratification}

\end{document}